\documentclass[twocolumn,english,prx,superscriptaddress]{revtex4-1}
\usepackage[utf8]{inputenc}
\usepackage{xcolor}
\usepackage{bm}
\usepackage{float}
\usepackage{graphicx}
\usepackage{physics}
\usepackage{amsmath}
\usepackage{amssymb}
\usepackage{times}
\usepackage{enumitem}
\usepackage[colorlinks=true,citecolor=blue]{hyperref}

\begin{document}

\raggedbottom

\title{Probing non-equilibrium dissipative phase transitions with trapped-ion quantum simulators}

\author{Casey Haack}
\email{chaack@mines.edu}
\affiliation{Department of Physics, Colorado School of Mines, Golden, Colorado 80401, USA}

\author{Naushad Ahmad Kamar}
\affiliation{Department of Physics and Astronomy, Michigan State University, East Lansing, MI 48823, USA}

\author{Daniel Paz}
\affiliation{Department of Physics and Astronomy, Michigan State University, East Lansing, MI 48823, USA}

\author{Mohammad Maghrebi}
\affiliation{Department of Physics and Astronomy, Michigan State University, East Lansing, MI 48823, USA}

\author{Zhexuan Gong}
\email{gong@mines.edu}
\affiliation{Department of Physics, Colorado School of Mines, Golden, Colorado 80401, USA}
\affiliation{National Institute of Standards and Technology, Boulder, Colorado 80305, USA}

\begin{abstract}
Open quantum many-body systems with controllable dissipation can exhibit novel features in their dynamics and steady states. A paradigmatic example is the dissipative transverse field Ising model. It has been shown recently that the steady state of this model with all-to-all interactions is genuinely non-equilibrium near criticality,  exhibiting a modified time-reversal symmetry and violating the fluctuation-dissipation theorem. Experimental study of such non-equilibrium steady-state phase transitions is however lacking. Here we propose realistic experimental setups and measurement schemes for current trapped-ion quantum simulators to demonstrate this phase transition, where controllable dissipation is engineered via a continuous weak optical pumping laser. With extensive numerical calculations, we show that strong signatures of this dissipative phase transition and its non-equilibrium properties can be observed with a small system size across a wide range of system parameters. In addition, we show that the same signatures can also be seen if the dissipation is instead achieved via Floquet dynamics with periodic and probabilistic resetting of the spins. Dissipation engineered in this way may allow the simulation of more general types of driven-dissipative systems or facilitate the dissipative preparation of useful many-body entangled states. 

\end{abstract}

\maketitle

\newcommand{\avg}[1]{\langle #1 \rangle} 

\section{Introduction}
Quantum simulators are devices that allow for the study of the properties of quantum systems of interest in a controlled setting \cite{bloch_quantum_2012,georgescu_quantum_2014}. Because simulation of quantum mechanics by classical computers is untenable for even modestly sized systems, quantum simulators provide the best way to investigate novel quantum many-body physics. Among various platforms for quantum simulation \cite{zhang_superconducting_2023,yan_observation_2013,saffman_quantum_2010,kim_quantum_2010}, trapped ions have a unique advantage of simulating tunable long-range interacting quantum spin systems \cite{monroe_programmable_2021}. When interactions are sufficiently long-range, novel many-body phenomena can be observed, such as nonlinear lightcones \cite{Richerme2014,Jurcevic2014}, fast quantum information scrambling \cite{garttner_measuring_2017}, dynamical phase transitions \cite{Jurcevic2017,zhang_observation_2017,de2023non}, and continuous symmetry breaking in low dimensions \cite{chen_continuous_2023,feng_continuous_2022}.

However, so far most quantum simulation experiments, including those with trapped ions, focused on the simulation of closed quantum systems. The experimental simulation of open quantum systems with controllable dissipation is much less explored, despite the fact that these systems are often challenging to understand theoretically. One notable example is the simulation of a driven-dissipative quantum many-body system, where the competition between dissipation, interactions, and external drives can lead to a variety of steady-state phase transitions, many outside the paradigms of either classical or quantum phase transitions \cite{diehl_quantum_2008}. On the other hand, the ability to control dissipation in a quantum simulator also allows for dissipative quantum state engineering \cite{verstraete_quantum_2009,barreiro_open-system_2011,bermudez_dissipation-assisted_2013} or dissipative quantum error correction \cite{reiter_dissipative_2017}.

In this work, we show that state-of-art trapped-ion quantum simulators can be readily used to simulate driven-dissipative quantum spin systems and probe their non-equilibrium behaviors. As an example, we focus on the dissipative transverse-field Ising model (TFIM) with tunable long-range interactions \cite{overbeck_multicritical_2017,paz_driven-dissipative_2021,sierant_dissipative_2021}. First, we propose practical experimental setups to simulate this model with controllable dissipation for two commonly used hardware platforms: Yb$^+$ ions in a linear Paul trap and Be$^+$ in a Penning trap. We then show that a steady-state phase transition can be observed in practice for sufficiently long-ranged Ising interactions if the transverse field and the Ising interactions have the opposite sign.

Despite their inherently non-equilibrium dynamics, it has become increasingly clear that an effective thermal behavior emerges in a large variety of driven-dissipative phase transitions \cite{mitra_nonequilibrium_2006,wouters_absence_2006,torre_keldysh_2013,lee_unconventional_2013,sieberer_keldysh_2016,maghrebi_nonequilibrium_2016,owen_quantum_2018,chan_limit-cycle_2015,wilson_collective_2016,le_boite_steady-state_2013,foss-feig_emergent_2017}, although non-equilibrium or quantum behavior emerge upon fine tuning or more complex dynamics \cite{Sieberer13,Altman2015,Marino_2016,Young_2020,Marino_2022}. For example, a system in thermal equilibrium satisfies the principle of detailed balance, leading to a time reversal symmetry (TRS) as well as fluctuation-dissipation relations between thermal fluctuations and the causal response of the system \cite{biroli_slow_2016}. A driven-dissipative system, even at its steady state and close to its phase transition point, may be qualitatively different from a thermal equilibrium system \cite{zwanzig_nonequilibrium_2001,agarwal_open_1973,carmichael_detailed_1976}. In a recent work \cite{paz_time-reversal_2021}, we have shown that the steady-state phase transition of a dissipative TFIM with all-to-all interactions is genuinely non-equilibrium in that a modified TRS shows up in the steady state at the phase boundary, thus violating the fluctuation-dissipation theorem.

Here we propose a practical method to measure such a non-equilibrium steady-state phase transition using state-of-the-art trapped-ion quantum simulators and reveal that the above-mentioned modified TRS can be observed with a system size as small as $10$ ions and are robust against a slow decay of interactions over distances. In addition, we show that both the steady-state phase transition and its non-equilibrium feature can also be observed with a stroboscopic dissipation engineered via Floquet dynamics and probabilistic resetting of each ion qubit. Such Floquet dissipation was first introduced in Ref.\,\cite{sierant_dissipative_2021} as a means to study measurement-induced phase transitions. We show that this Floquet dissipation is in fact equivalent to a continuous dissipation if the Floquet period is sufficiently small, and our numerical calculations indicate that the two types of dissipation generate qualitatively similar steady state properties as long as the Floquet period is not too large. We also anticipate such Floquet engineering of dissipation to be more flexible and capable of realizing more non-trivial types of dissipation that could facilitate the experimental simulation of novel non-equilibrium phases of matter.

\section{Model and Experimental Setup}

As demonstrated by many trapped-ion quantum simulation experiments \cite{kim_quantum_2010,islam_onset_2011,britton_engineered_2012,Richerme2014,bohnet_quantum_2016,garttner_measuring_2017,zhang_observation_2017,Jurcevic2017}, one can use a trapped ion to encode a spin-1/2 (qubit) in two of its lowest energy levels and apply spin-dependent optical dipole forces via lasers to induce interactions among the spins \cite{monroe_programmable_2021}. With proper engineering of the lasers, we can simulate the dynamics of the following spin-1/2 TFIM:
\begin{equation}\label{hamiltonian}
    H = \textstyle \sum_{i<j}J_{ij}\sigma_i^x\sigma_j^x - \Delta \sum_i\sigma_i^z
\end{equation}
where the Ising interaction \{$J_{ij}$\} can be usually approximated by a power law decay: $J_{ij} \approx J_0 /r_{ij}^\alpha$ with $r_{ij}$ being the distance between the ions $i$ and $j$ while $0<\alpha<3$ \cite{britton_engineered_2012,islam_onset_2011}. The nearest-neighbor interaction strength achieved in most experiments is on the order of 0.1KHz (in angular frequency), while the achievable transverse field strength is at least an order of magnitude larger \cite{monroe_programmable_2021}. Typically, the Ising interactions are anti-ferromagnetic (i.e. $J_{ij}>0$) while the sign of the transverse field strength $\Delta$ can be arbitrary.

By varying $\Delta$, experiments have observed signatures of quantum phase transitions in the ground state of the above Hamiltonian \cite{kim_quantum_2010,islam_onset_2011,feng_continuous_2022} as well as dynamical phase transitions after quantum quenches \cite{Jurcevic2017,zhang_observation_2017,de2023non}. However, so far a phase transition in the steady state of this Hamiltonian in the presence of controlled dissipation has not been observed experimentally. Here we propose practical setups for the current trapped-ion experimental platforms to engineer dissipation and observe the dissipation phase transition associated with Eq.\,\eqref{hamiltonian}, together with its non-equilibrium features.

\begin{figure}
    \includegraphics[width=\columnwidth]{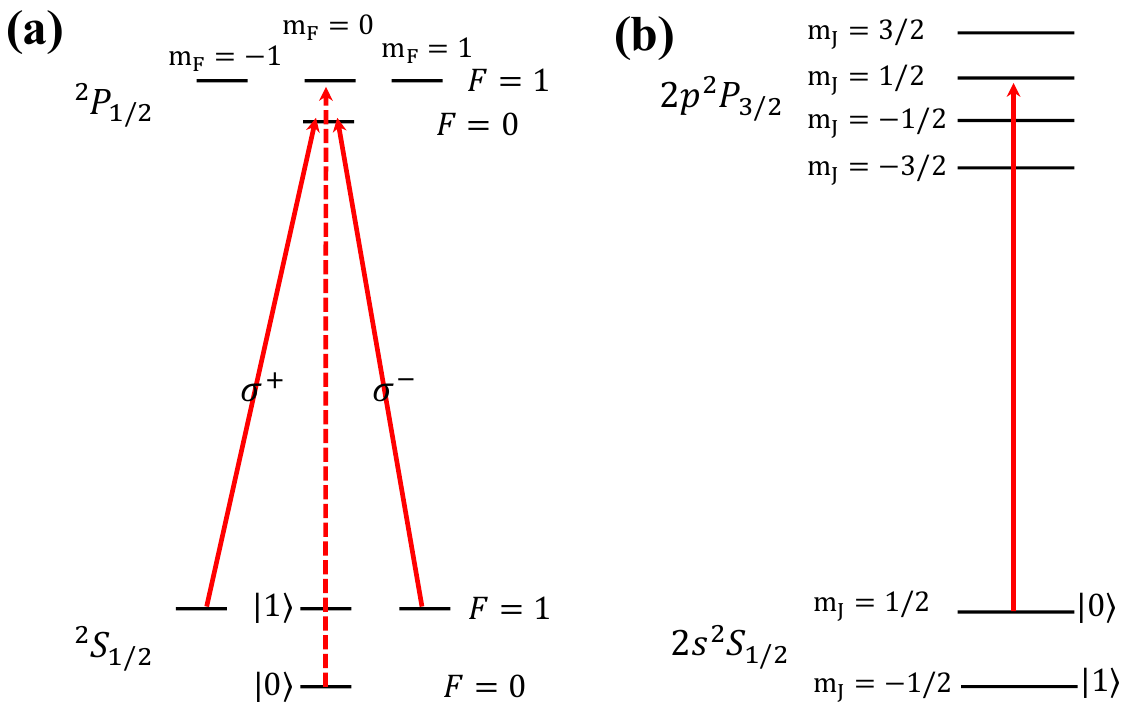}
    \caption{Relevant energy levels and proposed laser drives for achieving controllable dissipation for an $^{171}$Yb$^+$ ion in a linear Paul trap (a) and a $^9$Be$^+$ ion in a Penning trap (b). \label{fig:levels}}
\end{figure}

To be specific, we focus on two common experimental platforms used for simulating Eq.\,\eqref{hamiltonian}: $^{171}$Yb$^+$ ions in a linear Paul trap \cite{kim_quantum_2010,islam_onset_2011,Richerme2014,zhang_observation_2017} and $^9$Be$^+$ ions in a Penning trap \cite{britton_engineered_2012,bohnet_quantum_2016,garttner_measuring_2017}. The relevant energy levels for an ion in each platform are shown in Fig.\,\ref{fig:levels}. Our goal is to engineer continuous dissipation on each spin-1/2 on top of the coherent dynamics governed by the TFIM Hamiltonian in Eq.\,\eqref{hamiltonian}. We also require the dissipation rate to be tunable below and above the typical interaction strength in order to study a dissipative phase transition. 

For an $^{171}$Yb$^+$ ion, where the two qubit states $|0\rangle$ and $1\rangle$ are respectively encoded in the two hyperfine ground states $^2S_{1/2}\ket{F=0,m_F=0}$ and $^2S_{1/2}\ket{F=1,m_F=0}$, we can achieve controllable dissipation by applying a near-resonant drive that couples the $|0\rangle$ state to one or more (depending on the polarization of the drive) of $^2P_{1/2}\ket{F=1}$ states. These optically excited states will undergo spontaneous emission and decay to any of the four $^2S_{1/2}$ states (there is also a small probability for decaying to the $^2D_{3/2}$ states, but those $^2D_{3/2}$ states can be re-pumped back into the $^2S_{1/2}, F=1$ states \cite{Olmschenk2007}). We denote the probability for decaying to the $|0\rangle$ state as $\kappa$, which has been measured to be approximately $1/3$ \cite{Olmschenk2007}. By changing the detuning or strength of the drive, we can control the rate of this dissipative process, denoted by $\gamma$ below, to be on the order of the Ising interaction strength. 

However, the above process can cause the ion to end up in a non-qubit state (either of the $^2S_{1/2}\ket{F=1,m_f=\pm 1}$ states). This can be fixed by adding two additional drives resonant with the transitions between the $^2S_{1/2}\ket{F=1,m_f=\pm 1}$ states and the $^2P_{1/2}\ket{F=0,m_F=0}$ with $\sigma^+$ and $\sigma^-$ polarization respectively. We can set the Rabi frequencies of these two additional drives to be much larger (such as MHz) than the dissipation rate we want to achieve ($\sim$KHz), and any leakage outside the qubit subspace can thus be ignored. Note that all drive strengths should be much smaller than the hyperfine splittings ($\sim$GHz) so that we can safely ignore any off-resonant transitions.

For a $^9$Be$^+$ ion, if we ignore the small hyperfine coupling between the electron spin and the nuclear spin of the ion, the two qubit states $|0\rangle$ and $|1\rangle$ are directly encoded in the two lowest energy levels $^2S_{1/2}\ket{m_j=\pm 1/2}$ [see Fig.\,\ref{fig:levels}(b)]. In this case, to achieve controllable dissipation on each qubit, one just need to apply a laser near-resonant with the transition between the $|0\rangle$ state and either of the $^2S_{1/2}\ket{m_j=\pm 1/2}$ optically excited states. Such excited state will decay to either the $|0\rangle$ or the $|1\rangle$ state. We again denote the probability for decaying to the $|0\rangle$ state as $\kappa$, which has been measured to be approximately $1/2$ \cite{britton_engineered_2012}. The rate of this dissipative process, denoted by $\gamma$ can again be controlled by changing either the intensity or the detuning of the drive laser.

For both ion species, We can model the above engineered dissipation using two jump operators for a given ion: One is $|1\rangle\langle 0|=\sigma^-$, representing the process that an ion initially in $|0\rangle$ gets optically excited and ends up in $|1\rangle$. This process happens at rate $\gamma_e \equiv \gamma(1-\kappa)$. The other is $|0\rangle\langle 0|=\frac{I+\sigma_z}{2}$, corresponding to the process that an ion initially in $|0\rangle$ gets optically excited and then decays back to $|0\rangle$. This process takes place at rate $\gamma_d \equiv \gamma \kappa$. Note that an ion initially in $|1\rangle$ will not (or hardly) be affected by the above dissipative process. Assuming that the three drives in Fig.\,\ref{fig:levels}(a) are applied uniformly across all ions, we end up with the following master equation for the state $\rho$ of all spins:
\begin{align}\label{eq:mastereq}
      \frac{d\rho}{dt} = -i\comm{H}{\rho} & - \frac{\gamma_e}{2} \textstyle\sum_i \left( \sigma_i^+  \sigma_i^- \rho + \rho \sigma_i^+  \sigma_i^- - 2 \sigma_i^-\rho \sigma_i^+ \right) \nonumber \\ & - \frac{\gamma_d}{4} \textstyle\sum_i \left( \rho - \sigma_i^z\rho \sigma_i^z\right)
\end{align}
where $H$ is defined in Eq.\,\eqref{hamiltonian}. This master equation describes the dynamics of a driven-dissipative system, where the competition between the continuous dissipation and the TFIM can lead to interesting phase transitions in the steady state, as discussed in the next section for details. Note that for Yb$^+$ ions, we expect $\gamma_d \approx \gamma_e/2$, while for Be$^+$ ions we expect $\gamma_d \approx \gamma_e$.

\section{Steady-state Phases}

Let us first describe an intuitive picture of what steady-state phases one should expect from Eq.\,\eqref{eq:mastereq}. For pedagogical reasons, we start by assuming $J_{ij}<0$ for any $i,j$ in Eq.\,\eqref{hamiltonian}. In this case, the ground state of Eq.\,\eqref{hamiltonian} is in a ferromagnetic (FM) phase provided that the transverse field strength $|\Delta|$ is not too large. We now argue that the steady state of Eq.\,\eqref{eq:mastereq} should be in the FM phase if \{$J_{ij}$\} are sufficiently long-ranged, \{$|\Delta|, \gamma_e, \gamma_d$\} are sufficiently small, and that $\Delta<0$ \cite{maghrebi_nonequilibrium_2016,overbeck_multicritical_2017,paz_driven-dissipative_2021}. An intuitive explanation of this argument is as follows: if $|\Delta| \gg \max|J_{ij}|$, the steady state should be close to a polarized state with $\sigma_i^z=-1$, which is also close to the ground state of the Hamiltonian for $\Delta<0$. Therefore, the dissipation is effectively cooling the system to nearly zero temperature. If we gradually decrease $|\Delta|$, we still expect dissipation to cool the system, but now likely to a finite effective temperature. With appropriate values of $\Delta$ and the dissipation rates, the steady state can be similar to a low temperature state of the Hamiltonian. In 1D, the FM order will be destroyed at any finite temperature by thermal fluctuations if $\{J_{ij}\}$ are short ranged \cite{Peierls1936}, but could survive if $\{J_{ij}\}$ decay slower than a certain power law ($1/|i-j|^{\alpha}$ with $\alpha < 2$ \cite{Dutta2001}). In 2D, the FM order is always stable below a certain critical temperature.

If either $|\Delta|$ or $\gamma_e$ is too large, we expect the steady state to be in a paramagnetic (PM) phase where the spins are approximately aligned in the $-z$ direction. A FM-PM phase transition is therefore induced by changing either $\Delta$ or $\gamma_e$.

Experimentally, we usually have $J_{ij}>0$ \cite{monroe_programmable_2021}, so the above argument does not directly apply. However, as we shown in Appendix \ref{App:Hsign}, exact relations between the steady state properties of Eq.\,\eqref{eq:mastereq} with $H$ and those with $-H$ can be established. The FM phase can thus also be observed for $J_{ij}>0$ if $\Delta>0$. To quantify the FM order and detect the FM-PM phase transition, we can experimentally measure the following ferromagnetic order parameter $M_F$:
\begin{equation}\label{op}
    M_F \equiv \textstyle \sum_{i,j} \langle \sigma_i^x \sigma_j^x \rangle / N^2.
\end{equation}
Such measurement has been routinely performed in trapped-ion quantum simulation experiments \cite{Richerme2014,Jurcevic2014,feng_continuous_2022}. Individual qubit readout is also not required for measuring $M_F$ as one can just measure the collective spin $S^x \equiv \sum_i \sigma_i^x$.

If $J_{ij}$ and $\Delta$ are of opposite signs, the analysis above seems to suggest that the steady state could exhibit anti-ferromagnetic (AFM) order. In 1D, however, the AFM phase would not survive since the long-range interaction pattern $J_{ij} = J_0/|i-j|^{\alpha}$ is in fact frustrated and does not help stabilize the AFM order under thermal fluctuations. In order to observe the AFM phase in the steady state, one needs to use an alternating sign interaction pattern such as $J_{ij}= J_0 (-1)^{i-j}/ |i-j|^{\alpha}$, which may be realized with trapped ions by coupling the spins dominantly to the zigzag mode of the ion chain \cite{feng_continuous_2022}. For a 2D spin lattice realizable by Penning trap experiments \cite{britton_engineered_2012,bohnet_quantum_2016,garttner_measuring_2017}, the observation of AFM order is complicated by the presence of geometric frustration, and we leave such topic to future study. As a result, from now on we will focus on the study of the FM phase and FM-PM phase transition as they are easier for experimental observation.

We will now find the steady-state phase diagram of Eq.\,\eqref{eq:mastereq}. Let us start from the simplest case where we have uniform, all-to-all Ising interactions with $J_{ij}=J_0$ ($\alpha=0$). In this case, mean-field theory is expected to be accurate in the large $N$ limit, where quantum fluctuations shall vanish. However, the standard mean-field theory cannot calculate two-point correlation functions such as the experimentally measurable $\langle \sigma_i^x \sigma_j^x \rangle$. Here we instead use a cumulant expansion method up to the second order, where we assume that third-order correlations between the spins are negligible, i.e.:
\begin{equation}
    \avg{\sigma_i\sigma_j\sigma_k} \approx \avg{\sigma_i} \avg{\sigma_j\sigma_k} + \avg{\sigma_j} \avg{\sigma_i\sigma_k} + \avg{\sigma_k} \avg{\sigma_i\sigma_j}
\end{equation} 
where $\sigma_i$ is an arbitrary Pauli operator for the spin $i$. In addition, we set $\langle \sigma_i ^x \rangle = \langle \sigma_i ^y\rangle =0$ in the steady state due to the $Z_2$ symmetry of Eq.\,\eqref{eq:mastereq} upon flipping every spin's $x$ and $y$ components. Using the Heisenberg equation together with this cumulant expansion method, we end up with a system of matrix equations:
\begin{gather}
0 = 2\mathbf{W}-\gamma_e(1+Z)\\
0 = 4\Delta \mathbf{W}- (\gamma_e+\gamma_{d})\mathbf{X}\\
0 = 4Z\mathbf{JW} + 4\Delta \mathbf{W} + (\gamma_e+\gamma_{d})\mathbf{Y}\\
0= 2Z\mathbf{JX} - 2\Delta(\mathbf{X}-\mathbf{Y}) + (\gamma_e+\gamma_{d})\mathbf{W}
\end{gather}
where $Z\equiv \langle \sigma_i^z \rangle$ (the system has translational invariance so the choice of $i$ does not matter), and the matrices denoted by bold letters above are defined as:
\begin{gather}
    \mathbf{X}_{ij}=\avg{\sigma_i^x\sigma_j^x}(1-\delta_{ij})\\
    \mathbf{Y}_{ij}=\avg{\sigma_i^y\sigma_j^y}(1-\delta_{ij})\\
    \mathbf{W}_{ij}=\avg{\sigma_i^x\sigma_j^y}(1-\delta_{ij})\\
    \mathbf{J}_{ij}=J_0(1-\delta_{ij})
\end{gather}
We can reduce this system of matrix equations to:
\begin{equation}
    \left[8Z\mathbf{J}+8\Delta + \frac{(\gamma_e+\gamma_{d})^2}{2\Delta}\right]\mathbf{X} = 0
\end{equation}
Because of the transitional invariance in Eq.\,\eqref{eq:mastereq}, we can simultaneously diagonalize $\mathbf{J},\mathbf{X}$. As we focus on studying the FM phase, we can just consider the above matrix equation for the uniform eigenvector of $\mathbf{J},\mathbf{X}$, where the corresponding eigenvalue for $\mathbf{J}$ is equal to $J_{\text{total}}=(N-1)J_0$, which denotes the total Ising interaction energy per spin. This allows us to solve for the value of $Z$, and from that we can solve $\mathbf{X},\mathbf{Y},\mathbf{W}$, leading to:
\begin{equation}\label{eq:XXce}
    \avg{\sigma_i^x\sigma_j^x} = \frac{\gamma_e}{\gamma_e+\gamma_d}\frac{16\Delta(J_{\text{total}}-\Delta)-(\gamma_e+\gamma_d)^2}{4(J_{\text{total}})^2}
\end{equation}
This result coincides with the standard mean-field prediction of $\langle \sigma_i^x \rangle^2$ in the steady state. By setting the r.h.s. of the above equation to zero, we obtain
\begin{equation}\label{eq:boundary}
    \gamma_e+\gamma_d = 4\sqrt{\Delta(J_{\text{total}}-\Delta)}
\end{equation}
which represents a phase boundary predicted by the mean-field theory or this second-order cumulant expansion method.

To see whether this predicted phase boundary is accurate, we perform exact numerical calculations of $\avg{\sigma_i^x\sigma_j^x} \approx M_F$ in the steady state of Eq.\,\eqref{eq:mastereq} for $\alpha=0$. To make sure the Hamiltonian [Eq.\,\eqref{hamiltonian}] is extensive in energy, we normalize the interaction strength $J_{ij}$ such that $J_{\text{total}}=1$, meaning that $J_{ij}=J_0=1/(N-1)$ in this case. And for simplicity, we set $\gamma_e=\gamma_d=\gamma$, corresponding to an equal probability for the optically excited state to decay to $|0\rangle$ versus $|1\rangle$.

\begin{figure*}[t]
\centering
\includegraphics[width=0.33\textwidth]{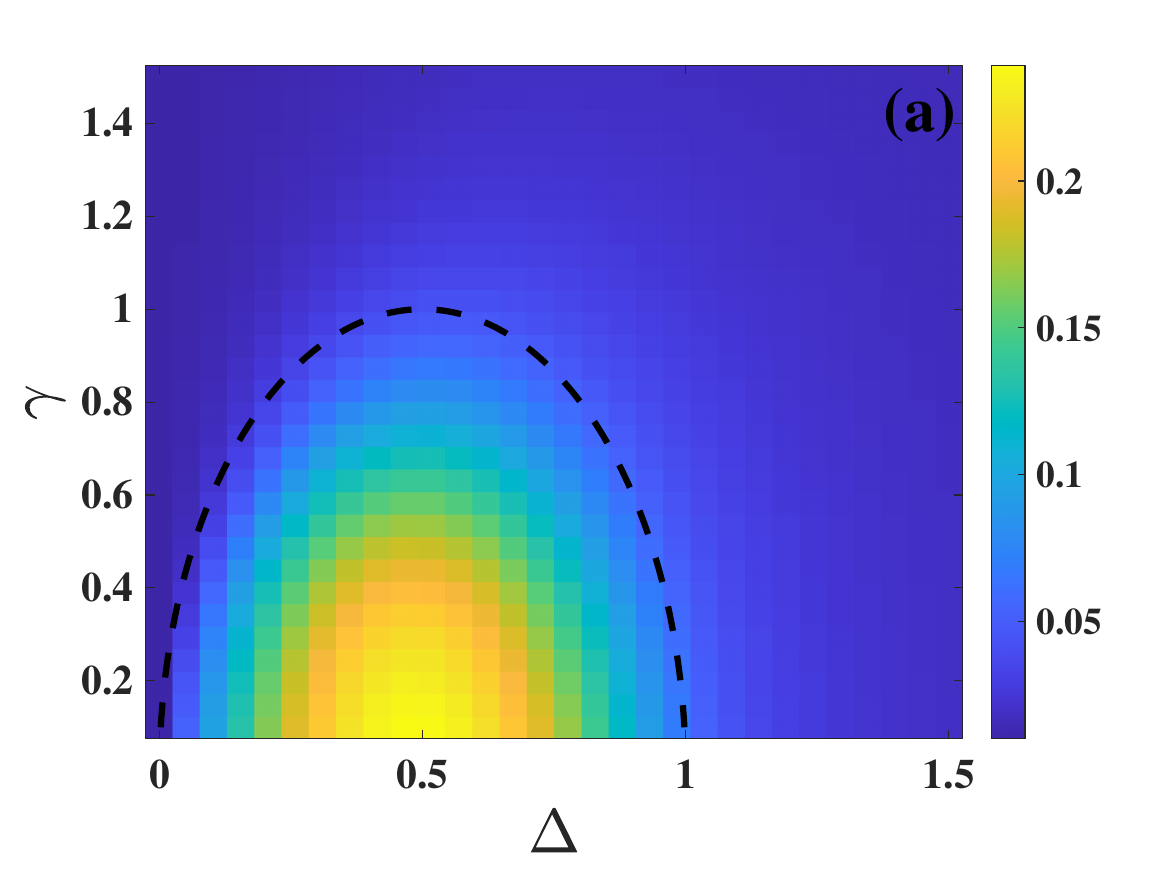}
\includegraphics[width=0.33\textwidth]{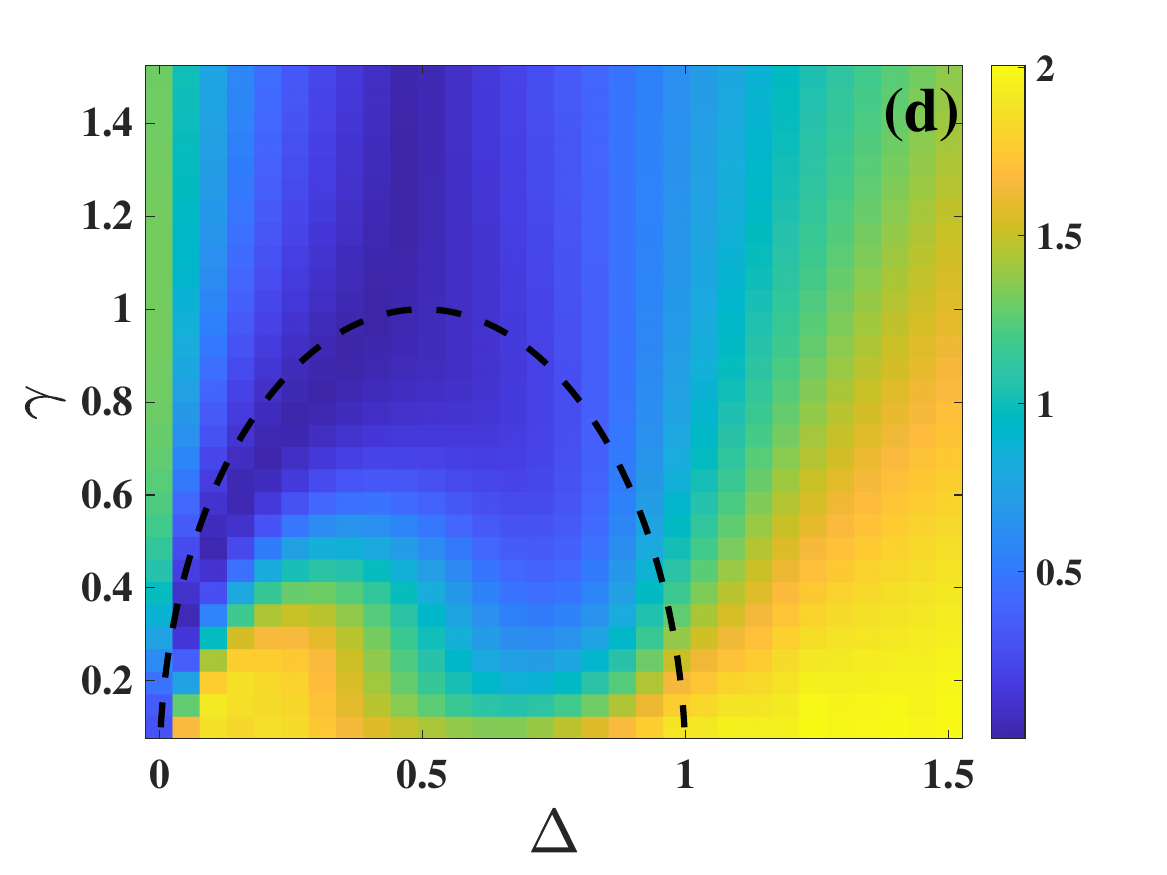}
\includegraphics[width=0.33\textwidth]{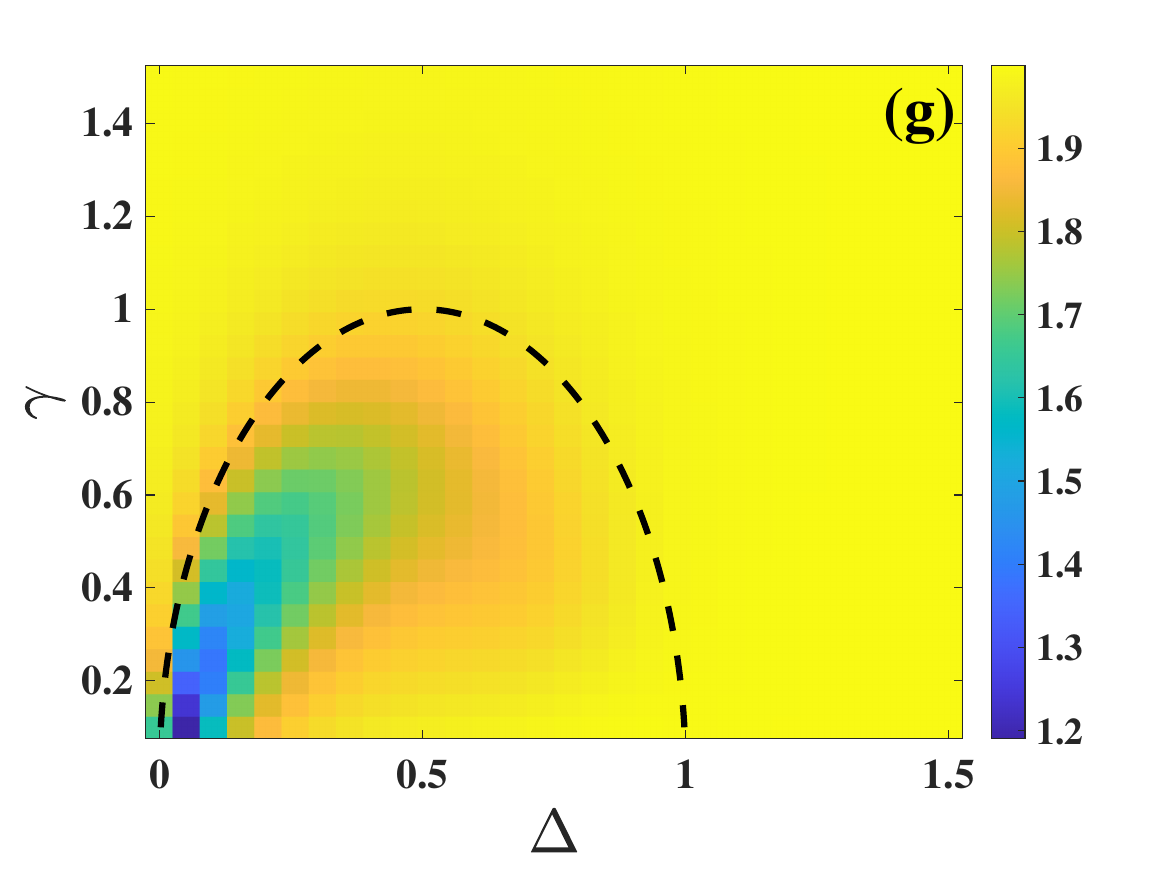}\\
\includegraphics[width=0.33\textwidth]{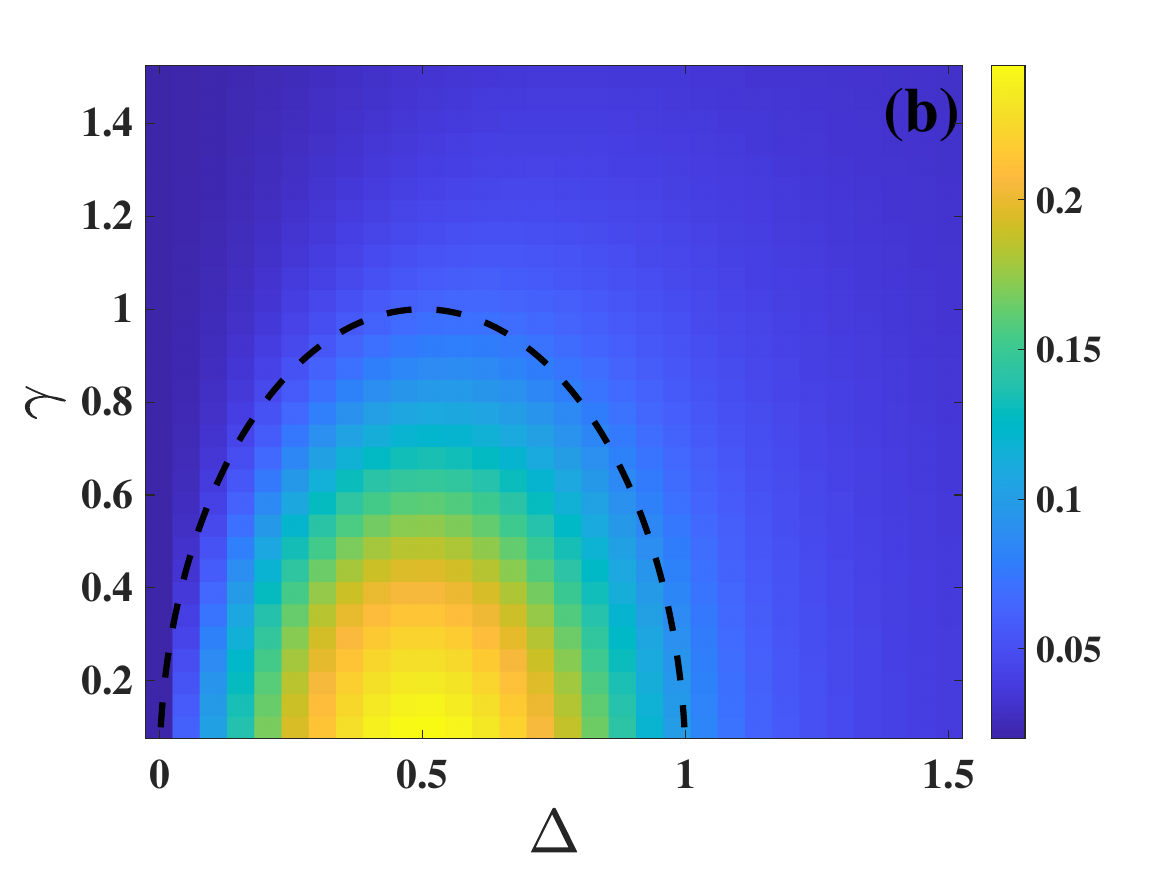}
\includegraphics[width=0.33\textwidth]{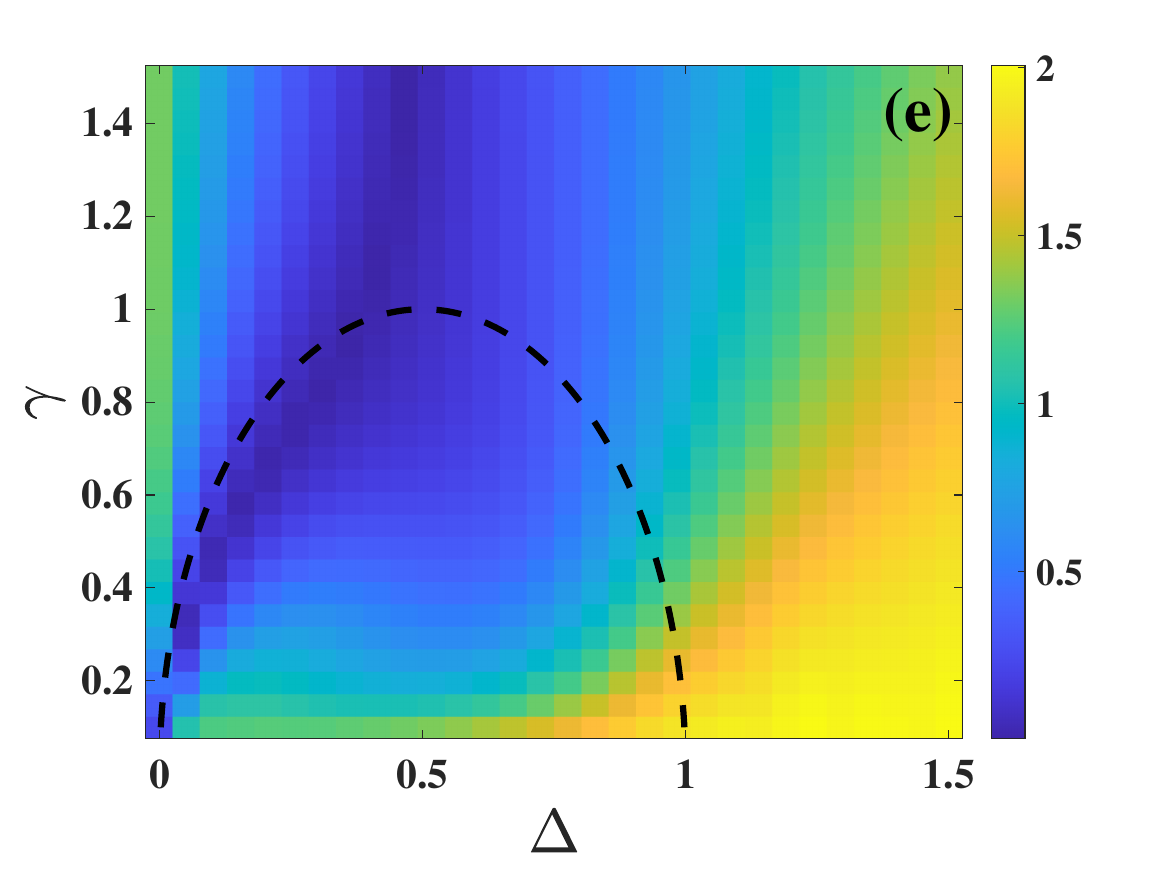}
\includegraphics[width=0.33\textwidth]{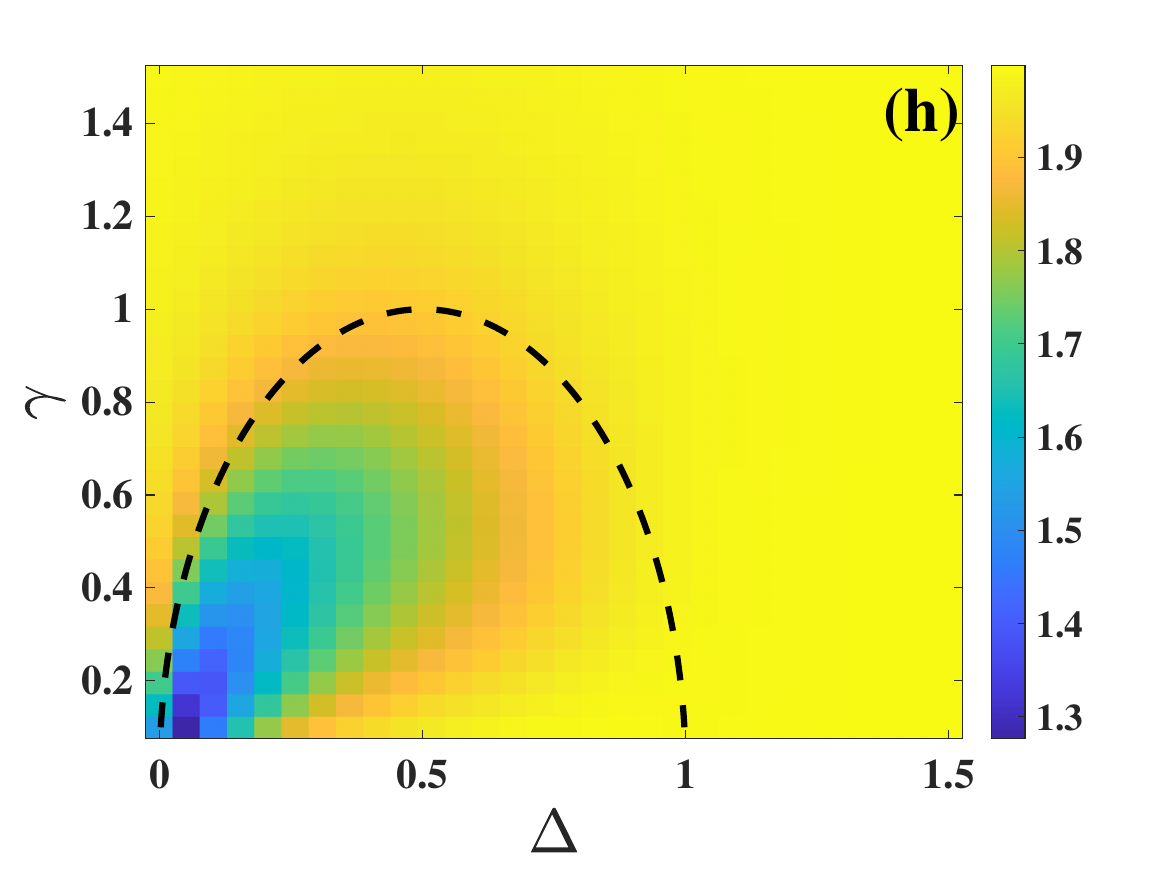}\\
\includegraphics[width=0.33\textwidth]{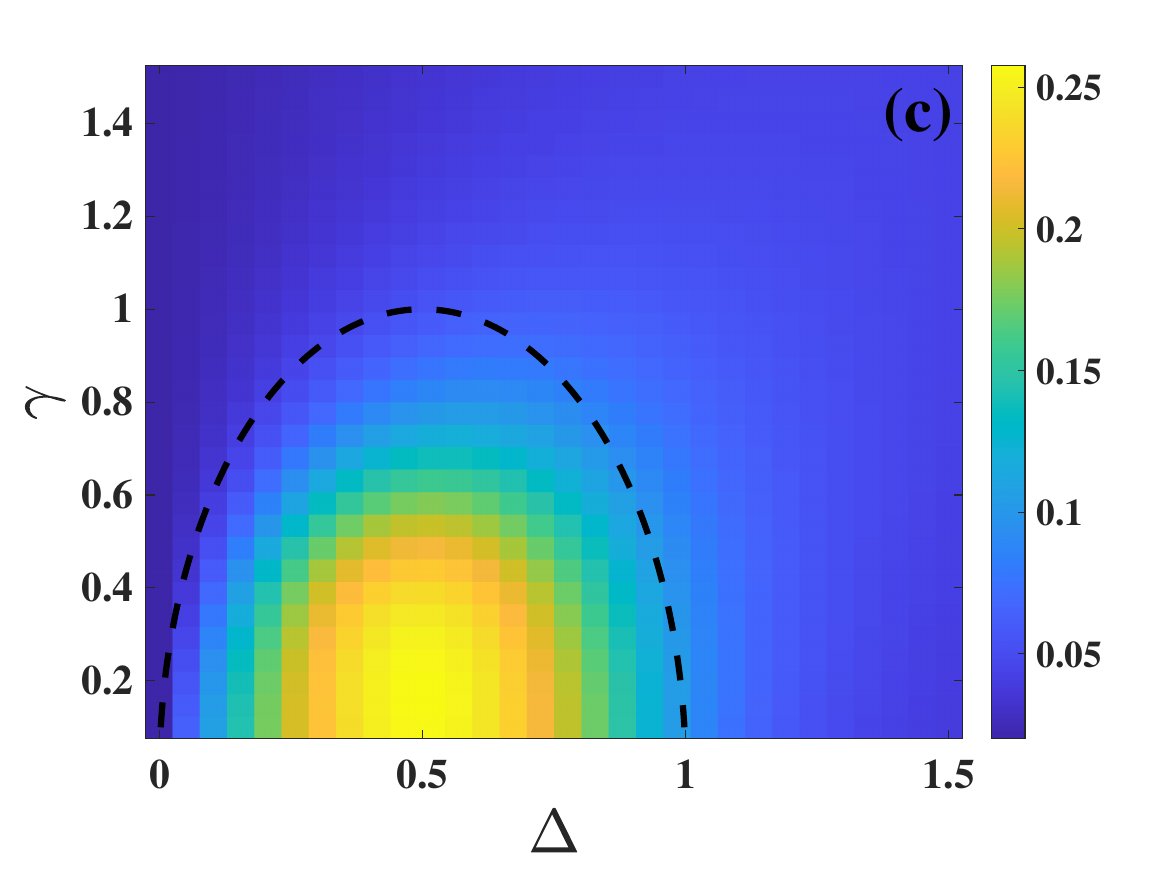}
\includegraphics[width=0.33\textwidth]{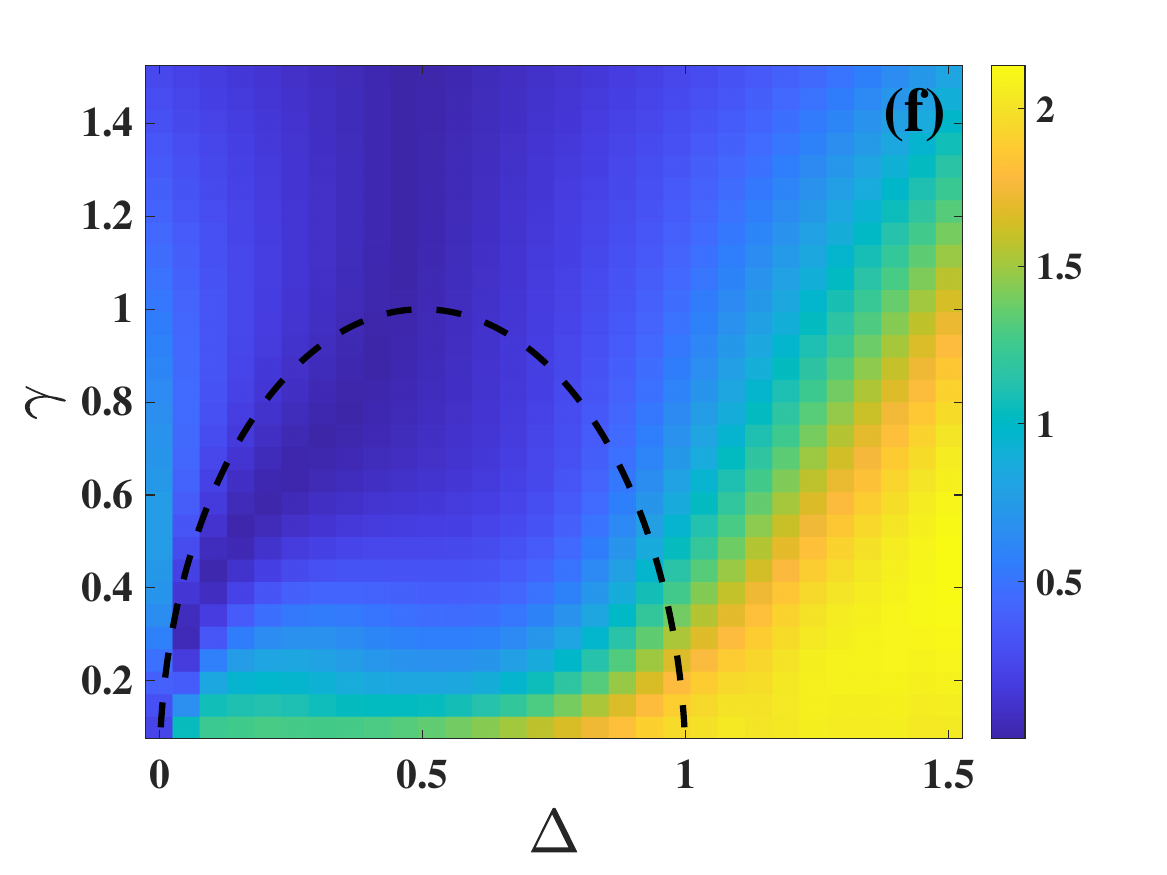}
\includegraphics[width=0.33\textwidth]{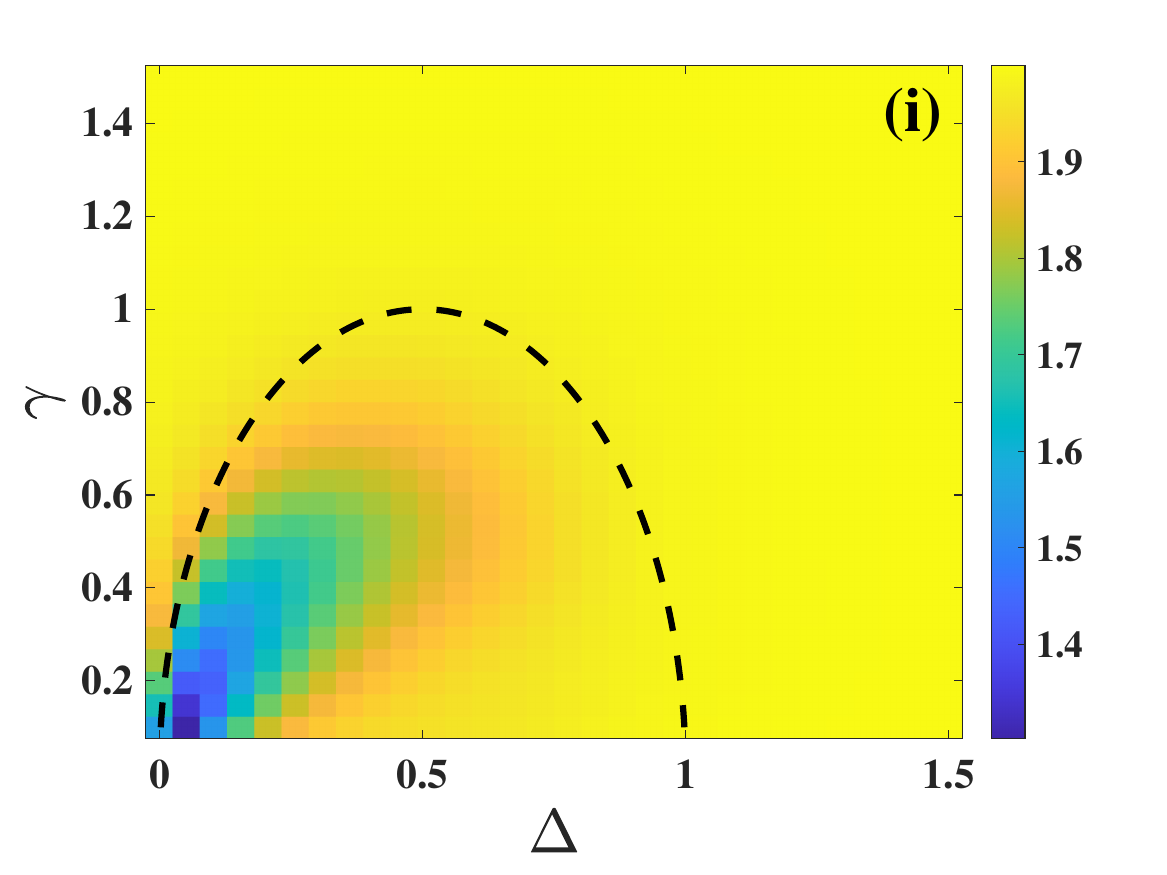}

\caption{ (a-c) Steady-state phase diagrams showing the FM order parameter $M_F$ [Eq.\,\eqref{op}] as a function of the transverse field strength $\Delta$ and dissipation rate $\gamma$ for (a) $N=100$ and (b) $N=50$ with continuous dissipation (Eq.\,\eqref{eq:mastereq} with $\gamma_e=\gamma_d=\gamma$), and (c) $N=50$ with Floquet dissipation (Eq.\,\eqref{Floquetevolve} with $\tau=0.5$ and $p=\gamma \tau$). The black dashed lines indicate the phase boundary predicted by Eq.\,\eqref{eq:boundary} using the second-order cumulant expansion method. (d-f) Similar to (a-c) but the value of $||C||$ is plotted instead of $M_F$. (g-i) Similar to (a-c) but the value of $||\chi||$ is plotted instead. $\alpha=0$ is assumed for this entire figure.} 
\label{fig:a0N100}
\end{figure*}

By utilizing the symmetry of Eq.\,\eqref{eq:mastereq} upon the permutation of any spins for $\alpha=0$, we can perform numerical simulation of Eq.\,\eqref{eq:mastereq} for $N=100$ spins easily with a high performance computer. Specifically, we choose a permutationally symmetric basis for the any permutationally invariant operators \cite{xu2013,paz_driven-dissipative_2021}, such that the density operator $\rho(t)$ in Eq.\,\eqref{eq:mastereq} can be expressed as a vector $\vec{\rho}(t)$ containing only $D=O(N^3)$ components, a drastic reduction from the $O(4^N)$ components in the general case. Both the Hamiltonian and the dissipative terms Eq.\,\eqref{eq:mastereq} can be written in this permutationally symmetric basis \cite{paz_driven-dissipative_2021} such that we can construct a Liouvillian $\mathcal{L}$ as a square matrix of dimension $D$ to rewrite Eq.\,\eqref{eq:mastereq} into:
\begin{equation}\label{Liouv}
    \frac{d\vec{\rho}(t)}{dt}= \mathcal{L} \vec{\rho}(t).
\end{equation}
The steady state $\vec{\rho}_s$ is then numerically obtained by finding the zero eigenvector of $\mathcal{L}$ via a shifted inverse power method. We then calculate the FM order parameter $M_F$ [Eq.\,\eqref{op}] using $\vec{\rho}_s$ and plot it in Fig.\,\ref{fig:a0N100}(a) as a function of both $\gamma$ and $\Delta$, which constitutes our numerical steady-state phase diagram.

For comparison, in Fig.\,\ref{fig:a0N100}(a) we have also shown the mean-field phase boundary obtained by setting $\gamma_e=\gamma_d=\gamma$ and $J_{\text{total}}=1$ in Eq.\,\eqref{eq:boundary}. One can see that the shape of this phase boundary agrees well with the contours of the numerically calculated values of $M_F$. Finite size effects on the phase boundary are not strong, as shown by the comparison between the $N=100$ [Fig.\,\ref{fig:a0N100}(a)] and the $N=50$ [Fig.\,\ref{fig:a0N100}(b)] numerical phase diagrams.

We have also computed the numerical steady state phase diagrams for $\alpha>0$ in 1D. Due to the lack of permutation symmetry, here we are limited to a much smaller system size of $N=10$ spins for exact calculations. The Ising interactions now follow $J_{ij} = J_0/|i-j|^{\alpha}$ and we set $J_0$ such that the spatially averaged total interaction energy per spin $J_{\text{total}} \equiv \frac{1}{N} \sum_{j\ne i} J_{ij} =1$. Although the Ising interactions are not translationally invariant due to the open boundary condition, it can be shown that the uniform eigenvector is still approximately an eigenvector of the interaction matrix formed by $\{J_{ij}\}$, and therefore the mean-field phase boundary in Eq.\,\eqref{eq:boundary} still holds approximately with the above new definition of $J_{\text{total}}$. In fact, by setting $J_{\text{total}}=1$, the mean-field phase boundary is always given by $\gamma=2\sqrt{\Delta(1-\Delta)}$ regardless of the value of $\alpha$.

We numerically calculate the FM order parameter $M_F$ by evolving Eq.\,\eqref{eq:mastereq} via an ODE solver for a sufficiently long time ($T=100$). We notice that except when $\Delta$ or $\gamma$ values are very small, a much shorter evolution time of $T=20$ yields no noticeable difference in $M_F$, indicating that the evolved state should be very close to the steady state. Experimentally, $J_{\text{total}}$ is usually on the order of KHz (in angular frequencies) for $N\sim 10$ and thus we estimate the time needed to prepare the steady state to be a few milliseconds, which is within the coherence time of ion qubits for current experiments \cite{monroe_programmable_2021}.

\begin{figure*}[t]
\centering
\includegraphics[width=0.33\textwidth]{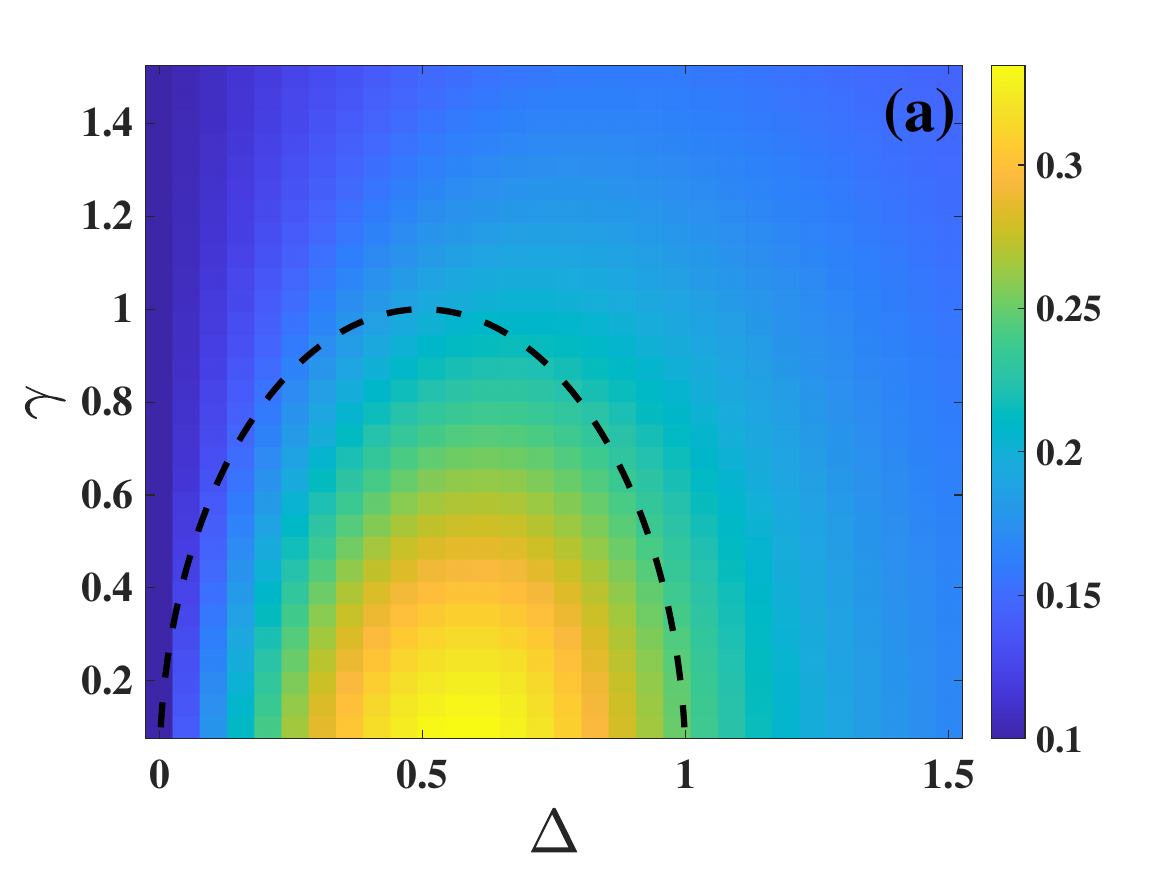}
\includegraphics[width=0.33\textwidth]{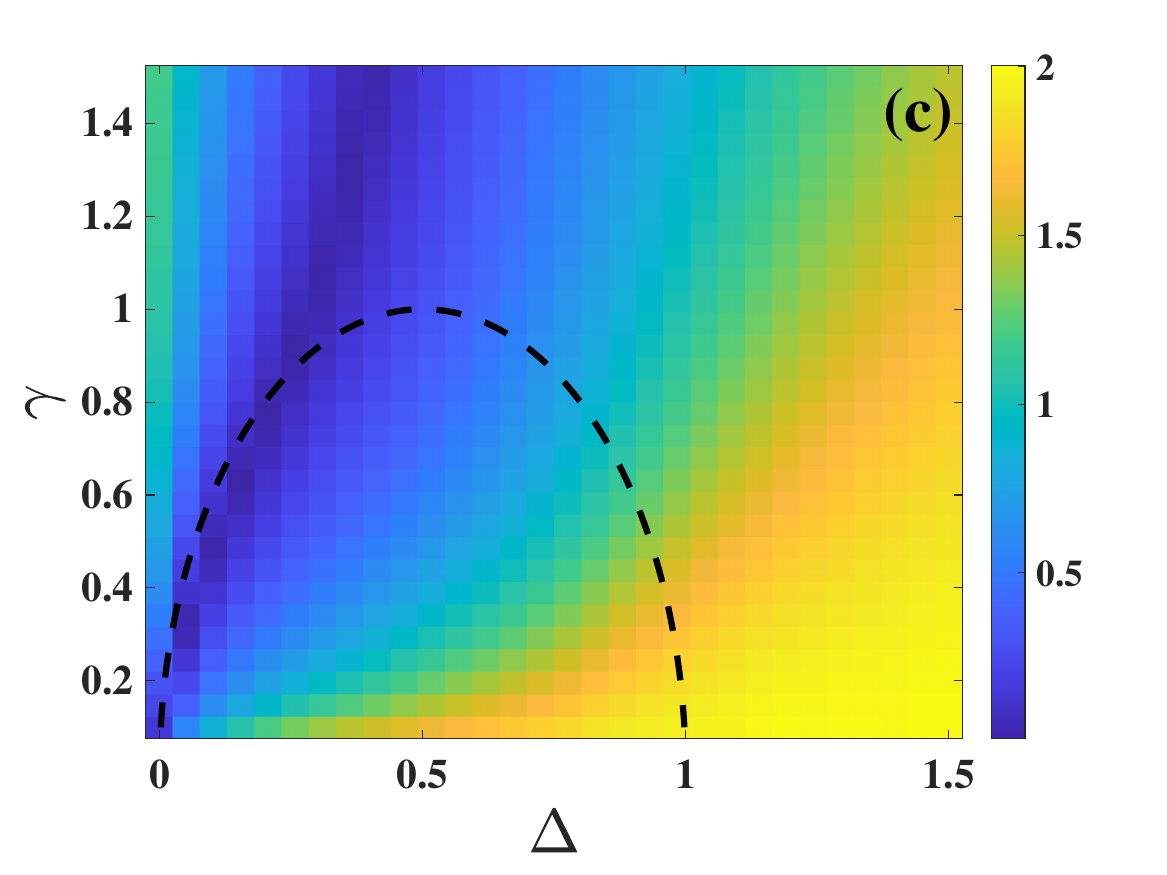}
\includegraphics[width=0.33\textwidth]{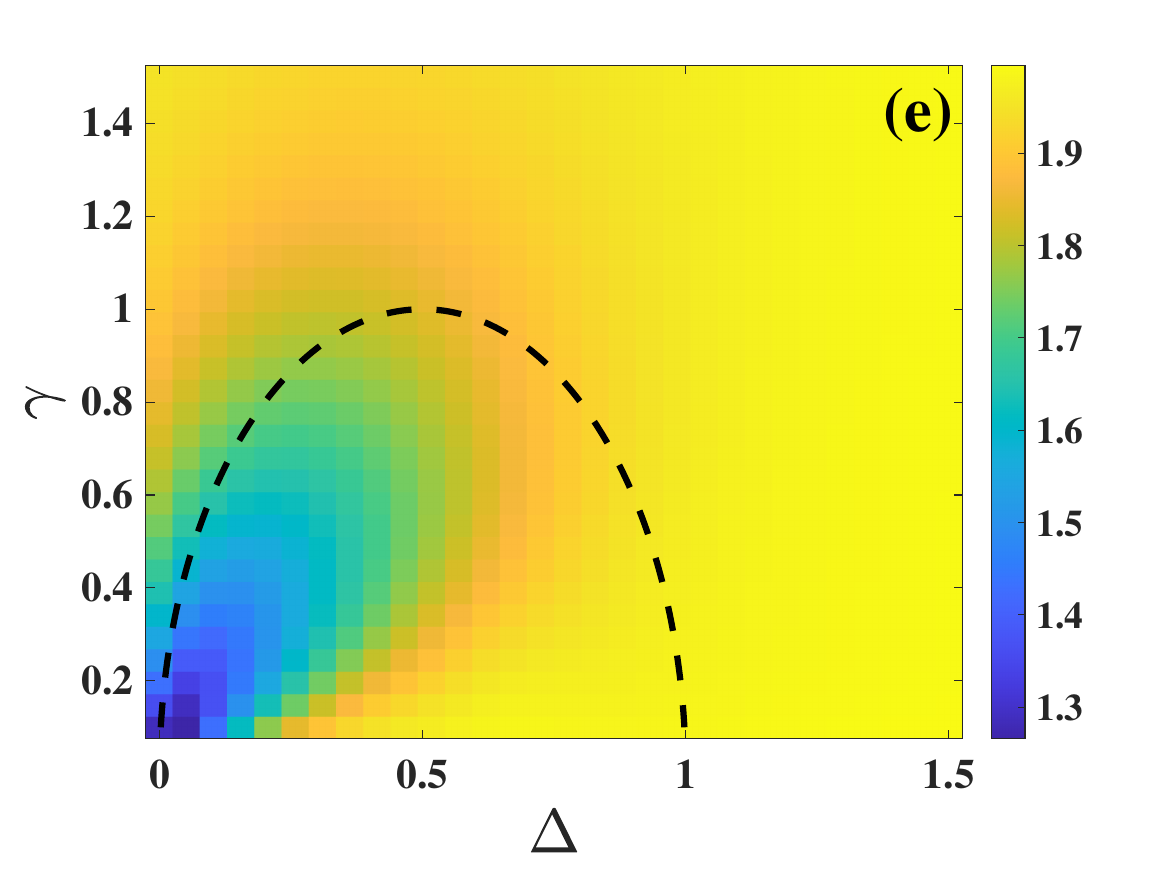}\\
\includegraphics[width=0.33\textwidth]{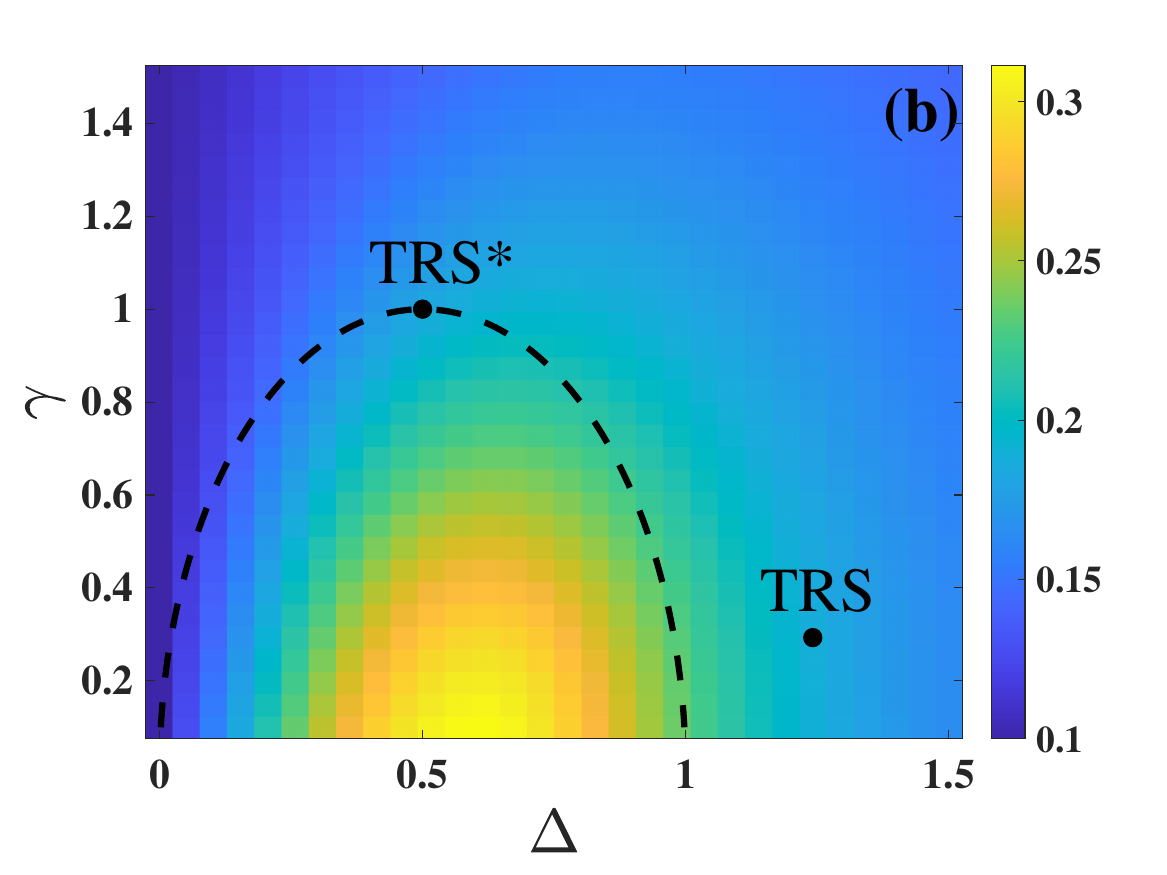}
\includegraphics[width=0.33\textwidth]{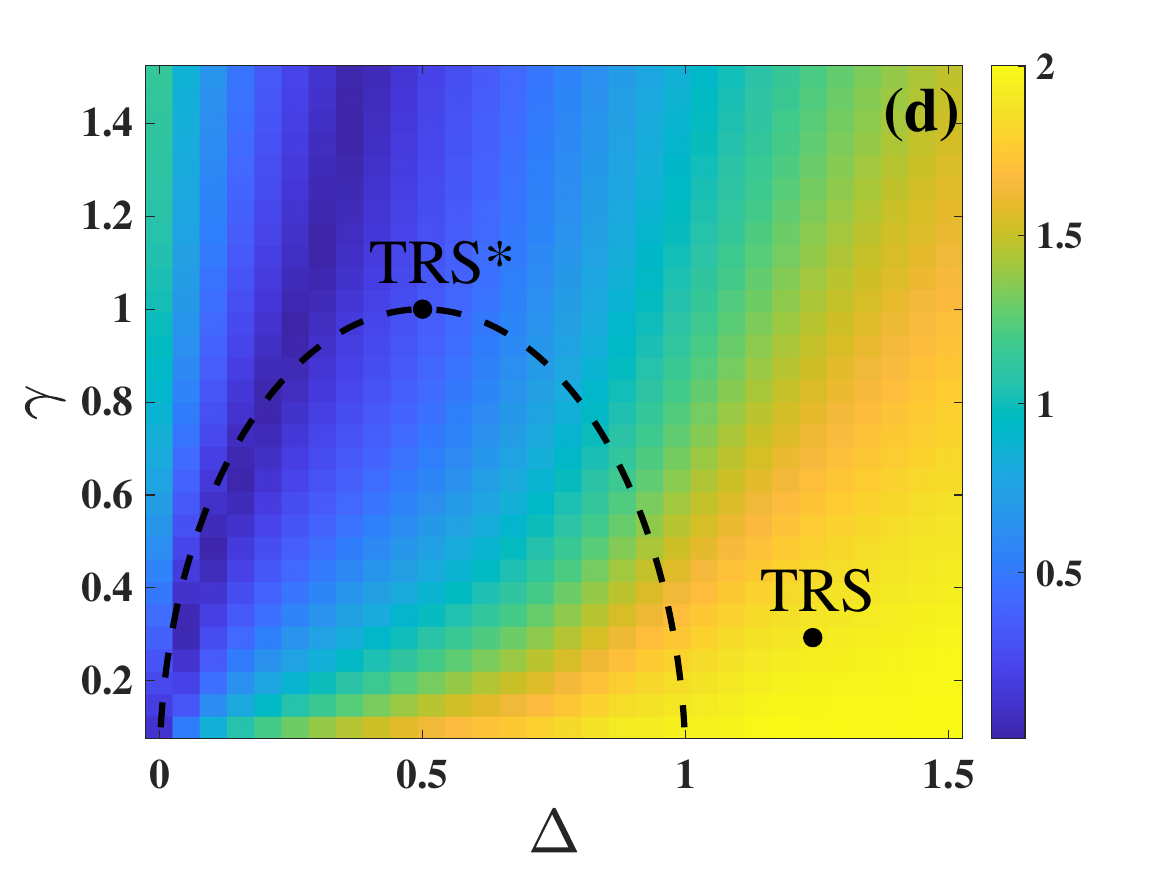}
\includegraphics[width=0.33\textwidth]{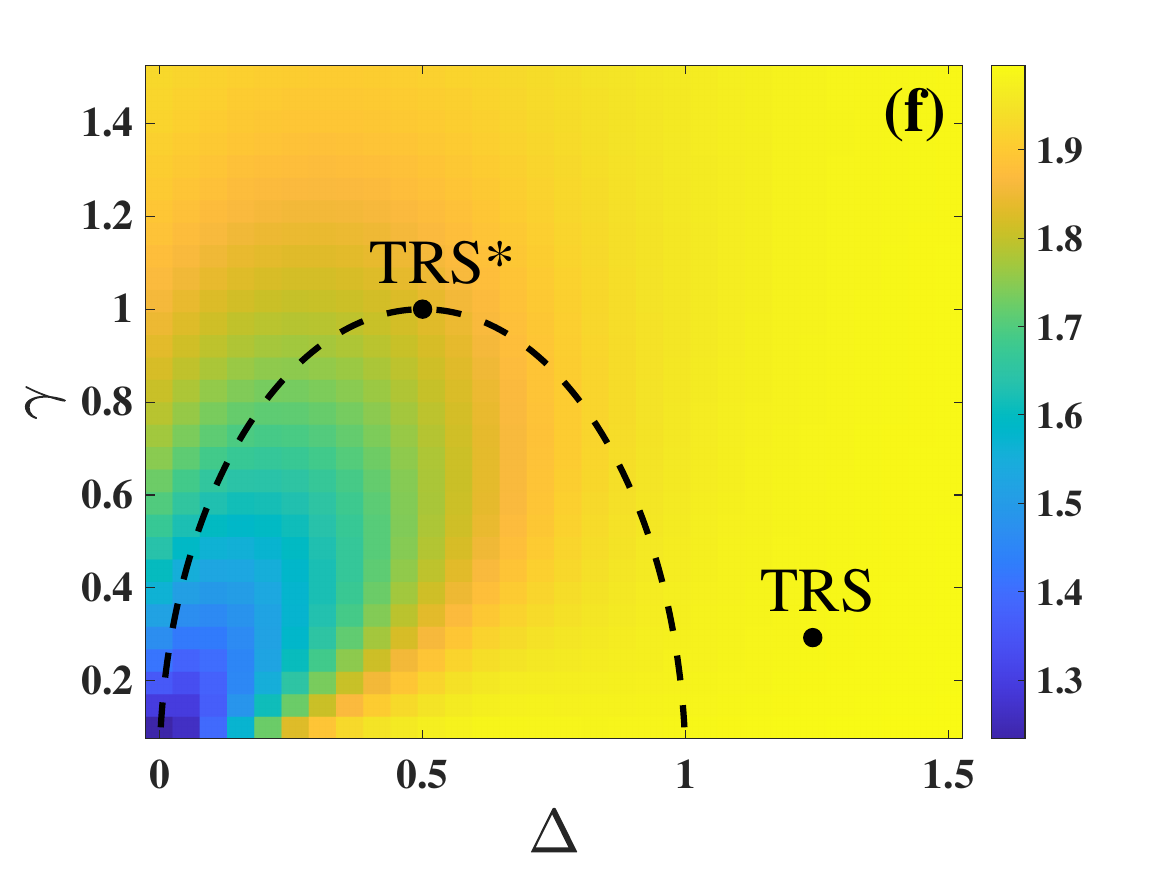}

\caption{The values of $M_F$ (a-b), $||C||$ (c-d), and $||\chi||$ (e-f) in the steady state of Eq.\,\eqref{eq:mastereq} as a function of $\Delta$ and $\gamma$ for $N=10$ with $\gamma_e=\gamma_d=\gamma$. The top row assumes $\alpha=0$ and the bottom row assumes $\alpha=1$. Exemplary points of TRS and TRS* are marked in the bottom row, corresponding to plots in Fig.\,\ref{fig:corrPlots}.}
\label{fig:N10}
\end{figure*}

As shown in Fig.\,\ref{fig:N10}(a), we see that the phase diagram indicated by $M_F$ for $\alpha=1$ and $N=10$ is qualitatively similar to that for a larger system [$N=100$ in Fig.\,\ref{fig:a0N100}(a) or $N=50$ Fig.\,\ref{fig:a0N100}(b)], despite of a less sharp phase boundary. Thus signatures of this dissipative phase transition can be experimentally observed with a system size and interaction pattern readily achievable for current ion-trap experiments \cite{monroe_programmable_2021}. The critical $\alpha$ above which the FM phase no longer survives (and the FM-PM phase transition disappears) is hard to determine. Ref.\,\cite{sierant_dissipative_2021} estimated $\alpha_c\approx 1.3$ with up to $N=24$ spins, but this estimation may not be accurate due to strong finite size effects. Nevertheless, we believe that the FM phase survives in the thermodynamic limit for all $0\le \alpha<1$ in 1D, and the mean-field phase boundary is exact in this regime due to strongly suppressed fluctuations. More precisely, one can show that the excitations beyond the collective sector are penalized by a finite gap \cite{paz_time-reversal_2021}.

\section{Non-equilibrium properties}
Steady-state phase transitions in driven-dissipative systems are often believed to be similar to phase transitions in thermal equilibrium (hence, classical phase transitions). However, this is not the case for the dissipative Ising model studied here. As shown in Ref.\,\cite{paz_time-reversal_2021}, genuine non-equilibrium behaviors of the steady-state phase transition can be found by observing the two-time correlation functions $\langle S^x(t) S^y(0)\rangle$ and $\langle S^y(t) S^x(0)\rangle$ in the steady state of Eq.\,\eqref{eq:mastereq} with all-to-all interactions,
where $S^{x,y} \equiv \sum_i \sigma_i^{x,y}$ are the collective spin operators. In particular, in thermal equilibrium due to time reversal symmetry (TRS) of the TFIM, we expect $\langle S^x(t) S^y(0)\rangle=-\langle S^y(t) S^x(0)\rangle$. On the other hand, TRS is explicitly violated by the dissipation in Eq.\,\eqref{eq:mastereq}, and the above two correlations are unrelated in general. However, upon approaching the dissipative phase transition, we recover an emergent, but modified, notion of time reversal symmetry (dubbed TRS*) where $\Re\langle S^x(t) S^y(0)\rangle=\Re\langle S^y(t) S^x(0)\rangle$ while $\Im\langle S^x(t) S^y(0)\rangle=-\Im\langle S^y(t) S^x(0)\rangle$ ($\Re$ and $\Im$ denote the real and imaginary parts of a complex number respectively), a behavior with no counterpart in equilibrium \cite{paz_time-reversal_2021}.

We attribute this non-equilibrium feature to the dissipative nature of the phase transition. Specifically, the order parameter in this model is a linear combination of $S^x$ and $S^y$: $S^\phi = \cos \phi S^x + \sin \phi S^y$. This should be contrasted against equilibrium where the order parameter is simply given by $S^x$ as TRS dictates that $\langle S^y\rangle=0$. Now, close to the dissipative phase transition, the contribution due to the order parameter $S^\phi$ is dominant. This implies that $\Re\langle S^x(t) S^y(0)\rangle$ and $\Re\langle S^y(t) S^x(0)\rangle$ are both dominated by $\Re \langle S^\phi(t)S^\phi(0)\rangle$ and are thus equal. The imaginary part of these two-time correlations require a different treatment since they probe the causal response of the system at time $t$ to an external field. A simple analysis can show that the system near the phase transition is only sensitive to an external field $h_\phi = -h_x \sin\phi + h_y \cos \phi$, where $h_x$ ($h_y$) denotes an external field along the $x$ ($y$) direction. It follows that $\partial \langle S^x(t)\rangle / \partial h_y $ is identical but opposite to $\partial \langle S^y(t)\rangle / \partial h_x $, hence $\Im\langle S^x(t) S^y(0)\rangle=-\Im\langle S^y(t) S^x(0)\rangle$. 

Here we go beyond the theoretical study in Ref.\,\cite{paz_time-reversal_2021} by providing experimentally practical methods to measure these two-time correlations and perform extensive numerical calculations to show that strong signatures of both TRS* and TRS can be observed in near-future ion-trap quantum simulation experiments with a small number of ions and practical interaction patterns. 

Typically, measuring two-time correlations is challenging and often requires the use of an ancilla qubit \cite{pedernales_efficient_2014} or non-Hermitian linear response \cite{Geier2022}. We provide a simple protocol to perform such measurement in current trapped-ion platforms that only requires individual addressing of the qubits, which has been routinely performed for ions in linear Paul traps \cite{monroe_programmable_2021} and also proposed for ions in Penning traps \cite{polloreno2022}. Our protocol is based on Ref.\,\cite{Uhrich2017} which was formulated for a closed quantum system only. We outline the protocol below and prove in Appendix \ref{App:Twotime} that this protocol works for a generic open quantum many-body system.

We first note that if the Hamiltonian in Eq.\,\eqref{hamiltonian} is translationally invariant (which should be approximately true for typical experiments), $\langle S^x(t) S^y(0)\rangle = N \langle S^x(t) \sigma_i^y(0)\rangle$ where $N$ is the number of qubits and $i=1,2,\cdots,N$. The steps for obtaining the real part of $\langle S^x(t) \sigma_i^y(0)\rangle$ are as follows:
\begin{enumerate}[itemsep=-1mm]
    \item Prepare the steady state by evolving an arbitrary initial state under Eq.\,\eqref{eq:mastereq} for long enough time.
    \item Projectively measure $\sigma_i^y$ and record the measurement outcome (either 1 or -1) via focused laser. All other qubits should be untouched during this measurement.
    \item Evolve the system under Eq.\,\eqref{eq:mastereq} for time $t$.
    \item Projectively measure $S^x$ and record the measurement outcome (ranging from $-N$ to $N$). 
    \item Repeat the above steps many times to obtain an average of the product of the above two measurement outcomes. 
\end{enumerate}
Appendix \ref{App:Twotime} shows that this average converges to $\Re \langle S^x(t) \sigma_i^y(0)\rangle$ as the number of repetitions increases. Note that in the second step above, to avoid excitations of other ion qubits from the measurement of the qubit $i$, one can perform measurement of the ion $i$ using a different cycling transition \cite{Olmschenk2007}, and a sympathetic cooling process can be performed to mitigate the heating of phonon modes due to this measurement. 

The steps for obtaining $\Im \langle S^x(t) \sigma_i^y(0)\rangle$ are:
\begin{enumerate}[itemsep=-1mm]
    \item Prepare the steady state by evolving an arbitrary initial state under Eq.\,\eqref{eq:mastereq} for long enough time.
    \item Rotate of the qubit $i$ around the $y$ axis by $\pi/2$.
    \item Evolve the system under Eq.\,\eqref{eq:mastereq} for time $t$.
    \item Projectively measure $S^x$ and record the outcome.
    \item Repeat the above steps many times to obtain the expectation value of $S^x$, denoted by $\langle S^x(t)\rangle_+$. 
\end{enumerate}
We then repeat the above procedure with a $-\pi/2$ rotation used in Step 2, and denote the result as $\langle S^x(t)\rangle_-$. As shown in Appendix \ref{App:Twotime}, $\Im \langle S^x(t) \sigma_i^y(0)\rangle = (\langle S^x(t)\rangle_+ - \langle S^x(t)\rangle_-)/\sqrt{2}$. The measurement of $\langle S^y(t) S^x(0)\rangle$ can be achieved similarly by simply exchanging the $x$ and $y$ directions for the measurement and rotations in the above protocols.

We now provide numerical evidences that by measuring $\langle S^x(t) S^y(0)\rangle$ and $\langle S^y(t) S^x(0)\rangle$, TRS* can be observed close to the phase transition while TRS can be observed away from the phase transition. For convenience, let us define
\begin{gather}\label{Cxy}
    C_{xy}(t) \equiv \frac{2}{N} \Re \langle S^x(t) S^y(0) \rangle = \frac{1}{N} \langle \{S^x(t), S^y(0)\} \rangle \\
    \chi_{xy}(t) \equiv \frac{2}{N} \Im \langle S^x(t) S^y(0) \rangle = \frac{1}{i N} \langle [S^x(t), S^y(0)] \rangle
\end{gather}
We similarly define $C_{yx}(t)$ and $\chi_{yx}(t)$ by 
swapping $x$ and $y$ in the above equation.

To quantify whether the steady state satisfies TRS or TRS*, here we define two new quantities $||C||,||\chi||$:
\begin{gather}
        ||C|| = ||\vec{C}_{xy}-\vec{C}_{yx}|| / ||\vec{C}_{xy}||\\
        ||\chi|| = ||\vec{\chi}_{xy}-\vec{\chi}_{yx}| / ||\vec{\chi}_{xy}||
\end{gather}
where $\vec{C}_{xy}$ denotes a vector formed by $C_{xy}(t)$ for $t\in [0,10]$ over $100$ time points (similarly for other vectors above), and $||\cdot ||$ denotes the Euclidean norm of a vector. Further increasing the range of $t$ or the number of time points do not lead to noticeably differences in $||C||$ or $||\chi||$ for our calculations. It's easy to see that if the steady state obeys TRS, we expect $||C||\approx 2$ and $||\chi||\approx 2$, while if the steady state obeys TRS*, we expect $||C||\approx 0$ and $||\chi||\approx 2$.

To see whether TRS* can be observed for a finite size system and for $\alpha, \gamma_d >0$ as in real experiments, here we perform exact numerical calculations of $\langle S^x(t) S^y(0) \rangle$ and $\langle S^y(t) S^x(0) \rangle$. First, we use the steady state $\rho_s$ of Eq.\,\eqref{eq:mastereq} found in Section III and then evaluate these two-time correlations using \cite{breuer_theory_2007}
\begin{equation}
    \avg{A(t)B(0)}= (\vec{A})^T e^{\mathcal{L}t} \vec{(B \rho_s)}
\end{equation}
where $A$ and $B$ and two arbitrary operators in the Hilbert space. The operators $A$ and $B \rho_s$ here are expressed as vectors in the Liouvillian space, while the Liouvillian $\mathcal{L}$, defined in Eq.\,\eqref{Liouv}, is expressed as a matrix.

Note that in the FM phase, the two-time correlations $C_{xy}(t)$ and $C_{yx}(t)$ approach nonzero values in the $t\rightarrow \infty$ limit, while we expect them to vanish everywhere else. Thus, to properly characterize the correlations, we subtract $C_{xy}(t)$ and $C_{yx}(t)$ by their respective values for the largest $t$ value ($t=10$) we calculated.

In Figs.\,\ref{fig:a0N100}(d-e), we show the value of $||C||$ across the phase diagram for $N=100$ and $N=50$ respectively. We have also calculated $||C||$ for a small system size of $N=10$ with $\alpha=0$ and $\alpha=1$, as shown in Figs.\,\ref{fig:N10}(c-d). We see that close to the phase boundary predicted by the mean-field theory, we indeed find that $||C||\approx 0$ and $||\chi||\approx 2$, consistent with TRS* and showing that the predictions in Ref.\,\cite{paz_time-reversal_2021} apply to more general types of dissipative Ising models (i.e. with dephasing and for small $\alpha>0$). Deviations from TRS* can be seen when approaching the right end of the phase boundary, which are likely due to finite size effects since smaller deviations are seen for larger system sizes. Remarkably, along the left half of the phase boundary, TRS* is not sensitive to finite size effects and can be clearly observed even for a small system size of $N=10$ for both $\alpha=0$ and $\alpha=1$. As a result, we expect the main signature of TRS* to be observable in near future experiments with a small chain of ions.

Next, we show the value of $||\chi||$ across the phase diagram in Fig.\,\ref{fig:a0N100}(g-i) and Fig.\,\ref{fig:N10}(e-f). We can see that for most part of the phase diagram, $||\chi||\approx 2$, meaning that $\chi_{xy}(t) = -\chi_{yx}(t)$, as expected for both TRS and TRS*.  For $\alpha=0$, the regions where $||\chi||$ deviates from 2 appears to shrink as the system size increases, consistent with field theory calculations in Ref.\,\cite{paz_time-reversal_2021} that  $||\chi||=2$ is expected in the thermodynamic limit across the entire PM phase.

We note that in the region with $\gamma \ll \Delta$ or $\gamma \ll 1$, $||C||\approx 2$ and $||\chi|| \approx 2$, meaning that TRS holds. This finding can be understood intuitively as follows: First, we note that $\langle S_x(t) S_y(0) \rangle^* = \langle S_y(0) S_x (t) \rangle$ upon Hermitian conjugation. Next, applying time reversal transformation (i.e. complex conjugation) on $\langle S_x(t) S_y(0) \rangle$ leads to $\langle S_x(t) S_y(0) \rangle^* = - \langle S_x(-t)S_y(0) \rangle = - \langle S_x(0) S_y(t) \rangle $. In the first equality, we have used the fact that the time evolution of $S_x$ is approximately given by that of the Hamiltonian when $\gamma\ll \Delta$ or $\gamma \ll 1$, hence $(S_x(t))^*= S_x(-t)$; in the second equality, we have used the fact that the steady state is time translationally invariant. As a result, we find that $C_{xy}(t) \approx - C_{yx}(t)$ as well as $\chi_{xy}(t) \approx - \chi_{yx}(t)$.

\begin{figure}
\centering
\includegraphics[width=0.49\textwidth]{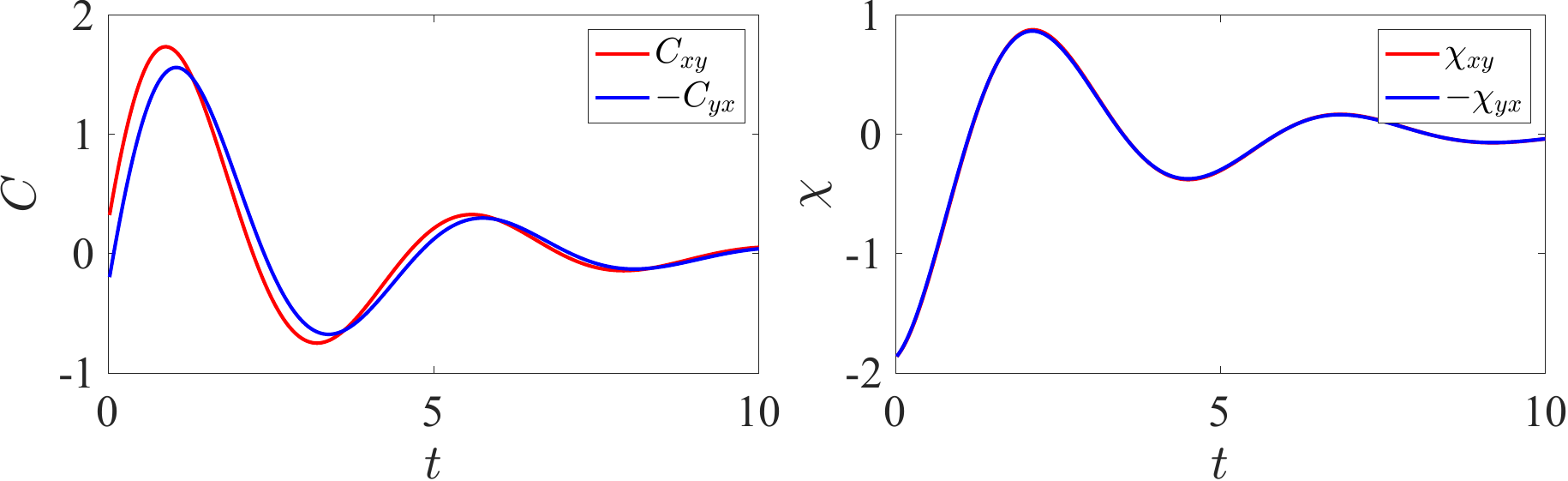}
\includegraphics[width=0.49\textwidth]{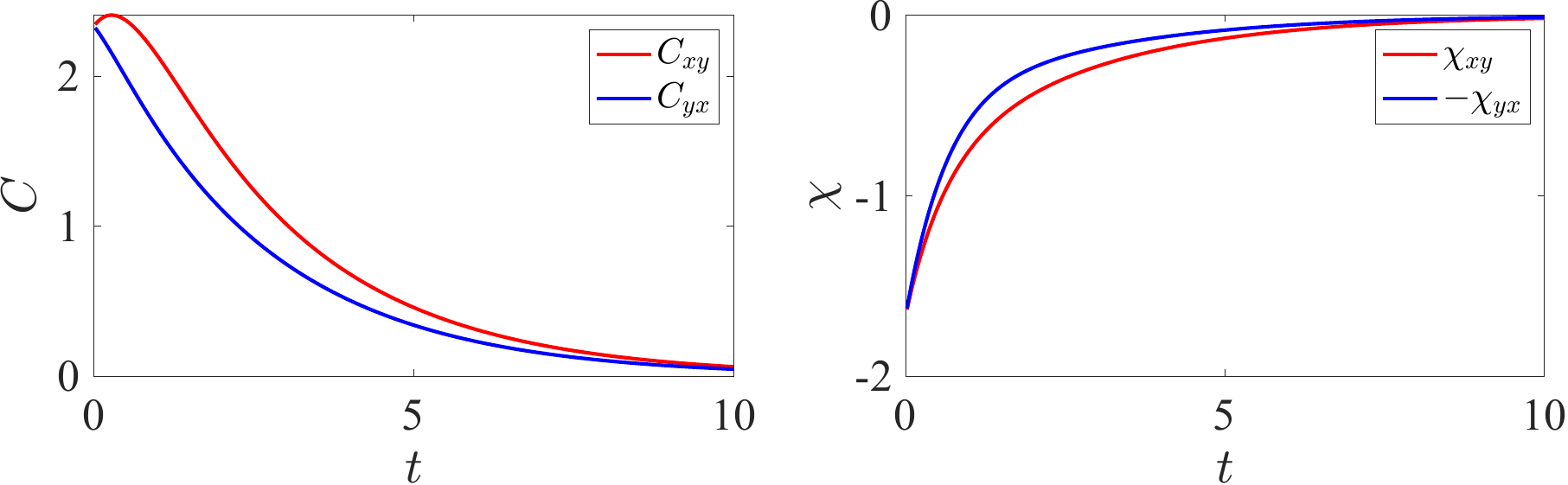}
\caption{Top Row: $C_{xy}(t)$, $C_{yx}(t)$, $\chi_{xy}(t)$ and $\chi_{yx}(t)$ for a point marked as TRS ($\Delta = 1.25,\gamma = 0.3$) in Fig.\,\ref{fig:N10}, with $\alpha=1,N=20$. Bottom Row: Same but for a point marked as TRS* ($\Delta=0.5,\gamma=1$) in Fig.\,\ref{fig:N10}, with 
$\alpha=1,N=20$.}
\label{fig:corrPlots}
\end{figure}

\begin{figure}
    \centering
    \includegraphics[width=0.35\textwidth]{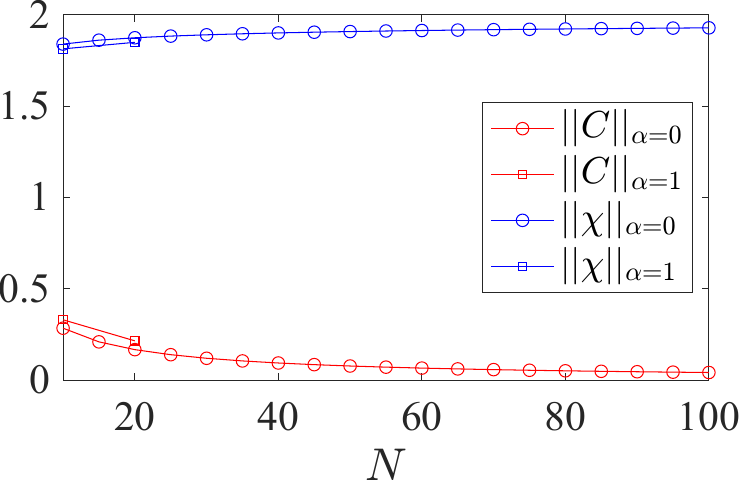}
    \caption{The values of $||C||$ and $||\chi||$ as a function of $N$ for $\alpha=0$ and $\alpha=1$ at the thermodynamic critical point of $\Delta=0.5,\gamma=1$.}
    \label{fig:normVsN}
\end{figure}

To show a direct comparison between TRS* and TRS in the steady state, we plot in Fig.\,\ref{fig:corrPlots} $C_{xy}(t)$, $C_{yx}(t)$, $\chi_{xy}(t)$, and $\chi_{yx}(t)$ for a point in Fig.\,\ref{fig:N10}(a) at the top of the mean-field phase boundary marked as TRS*, and a point deep in the PM phase with $\Delta \gg \gamma$ marked as TRS. We choose $N=20$ and $\alpha=1$ so that these curves can be possibly obtained in near-future experiments. It can be clearly seen that $C_{xy}(t)\approx C_{yx}(t)$ for the TRS* case and $C_{xy}(t)\approx - C_{yx}(t)$ for the TRS case, while for both cases $\chi_{xy}(t)\approx -\chi_{yx}(t)$. As we increase the system size, we expect both TRS* and TRS relations to become more and more accurate for the same values of $\Delta$ and $\gamma$. To see this, we plot the values of $||C||$ and $||\chi||$ at the aforementioned TRS* point with $\Delta=0.5$ and $\gamma=1$. As shown in Fig.\,\ref{fig:normVsN}, for $\alpha=0$ we find that $||C||\rightarrow 0$ and $||\chi||\rightarrow 2$ as $N$ increases from $10$ to $100$. It is therefore reasonable to expect TRS* to become exact in $N \rightarrow \infty$ limit. 

For $\alpha=1$, we calculate $||C||$ and $||\chi||$ up to $N=20$ by using a variational matrix product state (MPS) method \cite{Eduardo_MPS_steady_state_2015} to find the steady state and the approach in Ref.~\cite{Zalatel_Time-evolving_a_matrix_product_state_2015} to calculate two-time correlation functions. It is worth mentioning that we use a splitting approach to split the local Hilbert space dimension of the density matrix from $d^2$ to $d$ as in Refs.~\cite{kamar2023splitting, Shuo_pseudosite_MPS_2013}. For MPS calculations, we chose bond dimensions $50$ and $200$ for steady state and time evolution calculations, respectively. We find similar behaviors for $||C||$ and $||\chi||$ between $\alpha=1$ and $\alpha=0$ as $N$ is increased. We thus argue that TRS* should hold exactly at the FM-PM phase boundary for any $0\le \alpha < 1$. Again, this follows from the mean-field nature of the model for $\alpha<1$ where the dynamics is governed by a single mode representing the collective spin, while excitations to other (spin wave) modes are penalized by a finite gap \cite{paz_time-reversal_2021}.

\section{Dissipation via Floquet dynamics}

For experimental platforms where individual addressing of ions can be achieved, we may engineer effective dissipation for the TFIM in Eq.\,\eqref{hamiltonian} with Floquet dynamics \cite{sierant_dissipative_2021}. Instead of applying weak optical pumping lasers to achieve the continuous dissipation in Eq.\,\eqref{eq:mastereq}, we can periodically perform state preparation of each ion into the state $|1\rangle$ independently with a preset classical probability $p$ to achieve effective dissipation controlled by $p$. The advantage of this approach is that it requires no extra lasers and can be implemented directly in current experimental setups.

Since the state preparation process is usually much faster than the time scale of the spin Hamiltonian in typical trapped-ion experiments, we can treat the process as an instantaneous quantum map $\mathcal{E}$ described by:
\begin{align}\label{Floquetmapping}
    \mathcal{E}(\rho) &= \mathcal{E}_1\left(\mathcal{E}_2\left(\dots\mathcal{E}_N(\rho)\right)\right), \quad \mathcal{E}_i=\sum_{\mu=0,1,2}K_{i,\mu}\rho  K_{i,\mu}^{\dagger} \nonumber \\
    K_{i,0} &= \sqrt{p}\ketbra{1_i}{1_i}, K_{i,1}=\sqrt{p}\ketbra{1_i}{0_i}, K_{i,2}=\sqrt{1-p}I_i
\end{align}
Note that the order of the quantum maps \{$\mathcal{E}_i$\} is not important since the quantum maps on different ions mutually commute.

We implement a Floquet dynamics where a period $\tau$ of the system's evolution includes a coherent evolution of the TFIM Hamiltonian [Eq.\,\eqref{hamiltonian}] over time $\tau$ followed by the above dissipative quantum map $\mathcal{E}$. Denoting $\rho(n\tau)$ as the density operator of the system after $n$ periods, the equation of motion for the system is described by:
\begin{equation}\label{Floquetevolve}
    \rho\left[(n+1)\tau\right] = \mathcal{E}[e^{-iH\tau}\rho(n\tau)e^{iH\tau}]
\end{equation}

As we show in Appendix\,\ref{App:FvsC}, by using a Taylor series expansion in the time step $\tau$, in the $\tau\rightarrow 0$ limit the above Eq.\,\eqref{Floquetevolve} exactly reduces to the master equation [Eq.\,\eqref{eq:mastereq}] if we set $p=\gamma\tau$ and $\gamma_e=\gamma_d=\gamma$. 

To see whether the steady state obtained with this Floquet dissipation is actually similar to that obtained with the continuous dissipation for a finite $\tau$, we have performed exact numerical calculations of the steady states generated by Eq.\,\eqref{Floquetevolve}. Here we focus on the case of all-to-all Ising interactions ($\alpha=0$) where it's not hard to perform calculations for a system of $N=50$ qubits. In this case, we need to find a matrix $\mathcal{K}$ to rewrite the Floquet evolution Eq.\,\eqref{Floquetevolve} in the permutationally symmetric basis:
\begin{equation}
    \vec{\rho}[(n+1)\tau] = \mathcal{K} \vec{\rho}(n\tau)
\end{equation}
The construction of $\mathcal{K}$ is non-trivial and we refer the readers to Appendix \ref{App:Fmap} for technical details. The steady state density matrix is then obtained by finding the eigenvector of $\mathcal{K}$ with eigenvalue 1. We then compute the FM order parameter $M_F$ as well as the two-time correlations in the steady state.

For small $\tau$ values such as $\tau=0.1$ (in unit of $1/J_{\text{total}}$), we find almost no visible difference between the steady-state phase diagrams of the Floquet dissipation and those of the continuous dissipation, consistent with the perturbative analysis in Appendix \ref{App:FvsC}. Even for $\tau$ as large as $0.5/J_{\text{total}}$, as shown in Figs.\,\ref{fig:a0N100}(c,f,i), the differences between the two types of dissipation are small. This means the FM-PM phase transition and its non-equilibrium features can also be well observed with the Floquet dissipation. Experimentally, a large $\tau$ value is preferred since the Hamiltonian in Eq.\,\eqref{hamiltonian} only holds if the evolution time is much longer than the oscillation periods of the phonon modes used to mediate the Ising interactions \cite{monroe_programmable_2021}. This condition is well satisfied for a typical experiment with $N\sim 50$ Yb$^+$ ions even with $\tau=0.1/J_{\text{total}}$, where the phonon mode frequencies are around $5$MHz and $J_{\text{total}}\sim 5$KHz \cite{feng_continuous_2022}.

\section{Conclusion and Outlook}

In this paper, we propose novel experiments for current trapped-ion quantum simulation platforms to study steady-state phase transitions with controlled dissipation. We show that a dissipative Ising model with tunable long-range interactions can be realized in near-future experiments, where continuous dissipation can be engineered via lasers for commonly used ion species. By applying a transverse field with an opposite sign to the sign of the Ising interactions, the steady state can exhibit a FM phase for sufficiently long-range interactions. By changing either the transverse field strength or the dissipation rate, a FM-PM phase transition can be observed in the steady state by measuring the collective spin-spin correlations. No individual addressing is required for observing this steady-state phase transition.

In addition, we show that genuine non-equilibrium features of this dissipative phase transition can also be observed experimentally if individual addressing of the ion qubits can be achieved. We introduce a practical protocol for measuring a class of two-time correlations in the steady state relevant for the non-equilibrium features. Near the FM-PM phase transition, we find that such two-time correlation functions violate the time-reversal symmetry expected in a thermal equilibrium state. Remarkably, strong signatures of this non-equilibrium behavior can be observed with a small system size and does not require the interactions to be infinitely ranged as studied previously.

Finally, we investigated the differences between the TFIM with continuous dissipation and that with dissipation implemented via Floquet dynamics, first introduced in Ref.\,\cite{sierant_dissipative_2021}. We show analytically that the two models reduce to each other if the Floquet period approaches zero, and find numerically that even for relatively large Floquet periods, the steady state phases and the non-equilibrium features of the phase transition are very similar between the two models. This similarity is somewhat surprising as in this case the two models cannot be reduced to each other perturbatively. We are currently investigating if this similarity is expected for more general many-body Hamiltonians with a finite Floquet period.

An interesting future direction is to engineer more general types of dissipation via Floquet dynamics. By adding single-qubit rotations before and after the dissipative map in Eq.\,\eqref{Floquetmapping}, we can engineer almost arbitrary single-qubit jump operators. Moreover, if two-qubit gates are applied instead of single-qubit rotations, we can even engineer correlated jump operators that may be crucial in dissipate quantum state preparation \cite{verstraete_quantum_2009}. Such engineering can be easily achieved for a trapped-ion quantum computer.

We expect our work to open up a new thrust of quantum simulation experiments with trapped ions that focus on simulating driven-dissipative spin systems. For example, while here we focus on the dissipative Ising model as a paradigmatic example, trapped ions can be used to simulate XY or XXZ spin Hamiltonians \cite{Richerme2014,Jurcevic2014,feng_continuous_2022}. Rich driven-dissipative physics can be explored for such spin Hamiltonians with continuous symmetry \cite{maghrebi_nonequilibrium_2016,sieberer_keldysh_2016}. Together with the flexibility in engineering dissipation via Floquet dynamics, we believe a large variety of driven-dissipative systems can be studied in near-future quantum simulation experiments, many of which lack a thorough theoretical understanding as exact calculations are often limited to very small system sizes.

\begin{acknowledgments}
We thank Rajibul Islam, John Bollinger, and Guido Pagano for enlightening discussions related to this work and the HPC center at Colorado School of Mines for providing computational resources needed in carrying out this work. We acknowledge funding support from the NSF QIS program under the award number PHY-2112893. C. H. and Z.-X. G. are also supported by the W. M. Keck Foundation. N. K., D. P. and M. M. are also supported by AFOSR under the award number FA9550-20-1-0073 as well as the NSF CAREER Award DMR-2142866. 

\end{acknowledgments}

\appendix

\section{Relations between steady states of $H$ and $-H$}\label{App:Hsign}

As mentioned in Section III, here we show that the steady state properties of the dissipative TFIM [Eq.\,\eqref{eq:mastereq}] with the Hamiltonian $H$ in Eq.\,\eqref{hamiltonian} are related to those of the same model with $-H$. 

First, we note that both the Hamiltonian and the Lindblad operators are represented by real matrices in the computational basis. If we take the complex conjugate of both sides of Eq.\,\eqref{eq:mastereq}, we find that $\rho^*$ obeys the same master equation but with $H$ replaced by $-H$. Therefore, assuming we start from an arbitrary real initial density matrix, the steady state obtained with $H$ (denoted by $\rho_{H}$) should be the the complex conjugate of that with $-H$ (denoted by $\rho_{-H}$), i.e.
\begin{equation}\label{rhoH}
    \rho_H = \rho_{-H}^*
\end{equation}

As a result, the expectation values of any real operator $R$, which include $S^x, S^z$, and the ferromagnetic order parameter $M_F$, should be identical in $\rho_H$ and $\rho_{-H}$ since
\begin{equation}
   \tr(R\rho_H) = \tr(R\rho_{H})^* = \tr(R^*\rho_{-H}) = \tr(R\rho_{-H})
\end{equation}

On the other hand, the expectation value of any purely imaginary operator, such as $S^y$, in  $\rho_H$ is the opposite of that in $\rho_{-H}$.

Next, we investigate the relations between the two-time correlations such as $\avg{S^x(t)S^y}$ for $H$ and $-H$ evaluated in the respective steady state $\rho_H$ and $\rho_{-H}$. Since $S^x$ is a real operator, we have $S^x_H(t) = S^x_{-H}(t)^*$, similar to Eq.\,\eqref{rhoH}. Thus,
\begin{align}
    \tr [S^x_{H}(t) S^y \rho_H)] & = \tr [S^x_{-H}(t)^* S^y \rho_H)] \\ 
    & = \tr [S^x_{-H}(t) (S^y)^* \rho_H^*]^* \\ 
    & = - \tr [S^x_{-H}(t) S^y \rho_{-H}]^*
\end{align}
Therefore, we see that $C_{xy}(t)$ [defined in Eq.\,\eqref{Cxy}] flips its sign under the sign flip of the Hamiltonian, while $\chi_{xy}(t)$ remains the same.

The above relations should also hold for the Floquet dissipation described by Eq.\,\eqref{Floquetevolve}, since similar to the Lindblad operators in Eq.\,\eqref{eq:mastereq}, the Kraus operators in the dissipative map [Eq.\,\eqref{Floquetmapping}] are also real.

\section{Protocols for measuring two-time correlations} \label{App:Twotime}

The protocols mentioned in Section II for measuring real and imaginary parts of $\langle S^x(t)\sigma_i^y(0)\rangle $ was first introduced in Ref.\ \cite{Uhrich2017} for a closed quantum system. Here we show that the same protocol works for a generic open quantum system made of qubits and for measuring any two-time correlations in the form of $\langle A(t)\sigma_i(0)\rangle $, where $A(t)$ is an arbitrary Hermitian operator and $\sigma_i$ is an arbitrary Pauli operator (i.e. along any direction) for an arbitrary qubit $i$.

For pedagogical reasons, we will use the Schrodinger's picture from now on, where the two-time correlation $\avg{A(t)\sigma_i(0)}$ can be rewritten as
\begin{equation}\label{Spic}
    \avg{A(t)\sigma_i(0)} = \tr [ A \mathcal{E}_t (\sigma_i \rho_0)].
\end{equation}
Here $\rho_0$ is the density operator of the system at $t=0$, and $\mathcal{E}_t$ represents the quantum map that maps the state at $t=0$ to the state at time $t$ for any open quantum system. Using the Kraus representation of the quantum map $\mathcal{E}_t(O)=\sum_{\mu} \mathcal{K}_{\mu} O \mathcal{K}_{\mu}^{\dagger}$, we see that for any operator $O$ (not necessarily Hermitian)
\begin{equation}\label{Et}
    \mathcal{E}_t(O)^{\dagger} = \mathcal{E}_t(O^{\dagger}).
\end{equation}
Next, we define $P_+$ and $P_-$ as the projection operators onto the two eigenstates of $\sigma_i$ with eigenvalues $\pm 1$ respectively. Since $\sigma_i = P_+ - P_-$ and the identity operator $I=P_+ + P_-$, Eq.\,\eqref{Spic} becomes
\begin{align}
     \avg{A(t)\sigma_i(0)} & =   \tr [ A\mathcal{E}_t (P_+ \rho_0 P_+) ] -  \tr [ A \mathcal{E}_t (P_- \rho_0 P_-) ] \nonumber\\
    & +  \tr [ A\mathcal{E}_t (P_+ \rho_0 P_-) ]-  \tr [ A\mathcal{E}_t (P_- \rho_0 P_+)]
\end{align}
Using Eq.\,\eqref{Et}, we see that the last two terms in the above equation add to a purely imaginary number. Therefore,
\begin{equation}
    \Re \avg{A(t)\sigma_i(0)} = \tr [ A\mathcal{E}_t (P_+ \rho_0 P_+) ] -  \tr [ A \mathcal{E}_t (P_- \rho_0 P_-) ]
\end{equation}
Experimentally, this equation means that we can first measure $\sigma_i$ with all other qubits untouched, then evolve the resulting state (either $P_+ \rho_0 P_+$ or $P_- \rho_0 P_-$) for time $t$, and finally measure $A$. The expectation value of the product of the two measurement outcomes gives the real part of $\avg{A(t)\sigma_i(0)}$.

To measure the imaginary part of $\avg{A(t)\sigma_i(0)}$, we need to first apply a single-qubit unitary $e^{-i\pi \sigma_i/4}$ (i.e. a $\pi/2$ rotation) to the $i^{\text{th}}$ qubit. We then evolve the system for time $t$ and measure $A$ to get its expectation value, which is expressed as:
\begin{equation}
    \langle A(t) \rangle_{+} = \tr [\langle A \mathcal{E}_t (e^{-i\pi \sigma_i/2} \rho_0 e^{i\pi \sigma_i/2}) ]
\end{equation}
Since $e^{-i\pi \sigma_i/4} = \frac{1}{\sqrt{2}}(I - i \sigma_i)$, we obtain
\begin{align}
   2 \langle A(t) \rangle_{+} &=  \tr [A \mathcal{E}_t (\rho_0)] +  \tr [A \mathcal{E}_t (\sigma_i \rho_0 \sigma_i)] \nonumber \\
   &+ i \tr [A \mathcal{E}_t (\rho_0 \sigma_i)] - i  \tr [A \mathcal{E}_t (\sigma_i \rho_0)]
\end{align}
Next, we need to repeat the above measurement protocol with the single-qubit unitary $e^{i\pi \sigma_i/4}$ applied to the initial state instead. The expectation of $A(t)$ in the case is found to be:
\begin{align}
   2 \langle A(t) \rangle_{-} &=  \tr [A \mathcal{E}_t (\rho_0)] +  \tr [A \mathcal{E}_t (\sigma_i \rho_0 \sigma_i)] \nonumber \\
   &- i \tr [A \mathcal{E}_t (\rho_0 \sigma_i)] + i  \tr [A \mathcal{E}_t (\sigma_i \rho_0)]
\end{align}
Combining the above two equations and using Eq.\,\eqref{Et} yield:
\begin{gather}
   \langle A(t) \rangle_{+} - \langle A(t) \rangle_{-} =  i \tr [A \mathcal{E}_t (\rho_0 \sigma_i)] - i  \tr [A \mathcal{E}_t (\sigma_i \rho_0)] \nonumber \\
   = 2 \Im \tr [A \mathcal{E}_t (\sigma_i \rho_0)] = 2 \Im \avg{ A(t) \sigma_i(0)}
\end{gather}

Therefore, the imaginary part of $\avg{ A(t) \sigma_i(0)}$ is obtained by taking the difference of the measured expectation values of $A(t)$ for two different initial states, one with a $\pi/2$ rotation on the qubit $i$ and the other with a $-\pi/2$ rotation.

\section{Floquet dissipation in the small $\tau$ limit}\label{App:FvsC}

In this appendix, we show that the Floquet dynamics described by Eq.\,\eqref{Floquetevolve} are equivalent to the continuous dynamics governed by the master equation [Eq.\,\eqref{eq:mastereq}] up to the first order in the Floquet time step $\tau$ given $\gamma_e=\gamma_d=\gamma$ and $p=\gamma \tau$. To see this, we first expand Eq.\,\eqref{Floquetevolve} up to the first order in $\tau$. The coherent evolution part of Eq.\,\eqref{Floquetevolve} is expanded as
\begin{align}\label{rhotau}
    e^{-iH\tau}\rho(t) e^{iH\tau} = \rho(t) - i\tau\comm{H}{\rho(t)} +\mathcal{O}(\tau^2)
\end{align}
Next, we note that the dissipative map $\mathcal{E}_i$ defined in Eq.\,\eqref{Floquetmapping} can be explicitly written as
\begin{align} \label{Di}
\mathcal{E}_i(\rho) &= \rho + \gamma\tau \mathcal{D}_i(\rho)  \nonumber \\
\mathcal{D}_i (\rho) & \equiv \sigma_i^-\rho \sigma_i^+  + \frac{1-\sigma_i^z}{2} \rho \frac{1-\sigma_i^z}{2} - \rho
\end{align}
As a result, we find that
\begin{equation}
    \mathcal{E}(\rho)= \rho + \gamma \tau \sum_i \mathcal{D}_i(\rho) + O(\tau^2)
\end{equation}
Using Eq.\,\eqref{rhotau}, we end up with
\begin{align}
    \rho(t+\tau) &= \mathcal{E}( e^{-iH\tau}\rho(t) e^{iH\tau}) \nonumber \\
     & = \rho(t) - i \tau\comm{H}{\rho} + \gamma \tau \sum_i \mathcal{D}_i(\rho(t+\tau)) + O(\tau^2)
\end{align}
Therefore, up to the first order in $\tau$, we have an equation of motion under the Floquet dissipation:
\begin{equation}\label{Feq}
    \frac{d\rho}{dt}= -i \comm{H}{\rho}  + \gamma \sum_i \mathcal{D}_i(\rho)
\end{equation}
With a bit of algebra, we can rewrite $\mathcal{D}_i (\rho)$ in Eq.\,\eqref{Di} as:
\begin{equation}
    \mathcal{D}_i (\rho) = -\frac{1}{2}\left( \sigma_i^+  \sigma_i^- \rho + \rho \sigma_i^+  \sigma_i^- - 2 \sigma_i^-\rho \sigma_i^+ \right)  - \frac{1}{4} \left( \rho - \sigma_i^z\rho \sigma_i^z\right)
\end{equation}
Thus Eq.\,\eqref{Feq} reduces to the master equation [Eq.\,\eqref{eq:mastereq}] upon setting $\gamma_e=\gamma_d=\gamma$.

\section{Floquet dissipative map in a permutationally symmetric basis}\label{App:Fmap}

If both the coherent and dissipative dynamics of an $N$-qubit quantum system are invariant under any permutations of the qubits, we can find the steady state of the system using a permutationally symmetric basis for the density operator \cite{paz_driven-dissipative_2021,xu2013}:
\begin{widetext}
\begin{equation}
    \rho_{N_xN_yN_z} = \frac{1}{\sqrt{N!N_x!N_y!N_z!N_I}}\textstyle\sum_{P} \mathcal{P}_P
    \left(\sigma_1^x \cdots \sigma_{N_x}^x \sigma_{N_x+1}^y \dots \sigma_{N_x+N_y}^y \sigma_{N_x+N_y+1}^z \cdots 
    \sigma_{N_x+N_y+N_z}^z  I_{N_x+N_y+N_z+1}  \cdots I_{N} \right)
\end{equation}
\end{widetext}
where $N_x, N_y, N_z, N_I \in \{0,1,2,\cdots N\}$ represent the numbers of $\sigma^x, \sigma^y, \sigma^z$, and identity operators in the above basis state $\rho_{N_xN_yN_z}$ respectively. $N_I=N-N_x-N_y-N_z$, and $\mathcal{P}$ permutes the qubits according to the permutation $P$.

By assigning a single index to the basis state $\rho_{N_xN_yN_z}$, an arbitrary permutationally invariant density operator can be represented by a vector of dimension $O(N^3)$. The Liouvillian $\mathcal{L}$ in Eq.\,\eqref{Liouv} can be written as a matrix of the same dimension, and the explicit matrix elements of $\mathcal{L}$ for the master equation [Eq.\,\eqref{eq:mastereq}] can be found in Ref.\,\cite{paz_driven-dissipative_2021}.

For the Floquet dissipation, the Hamiltonian evolution part can be described by a similar Liouvillian matrix. The matrix elements for the dissipative map $\mathcal{E}$ [Eq.\,\eqref{Floquetmapping}] are harder to be found, and here we provide their analytical expressions:
\begin{widetext}
\begin{equation}\label{FloquetMappingPerm}
    \mathcal{E}(\rho_{N_xN_yN_z}) = (1-p)^{N_x+N_y+N_z}\textstyle\sum_{k=0}^{N_I}\binom{N_I}{k}(-p)^k\sqrt{\textstyle\prod_{i=1}^{k}\frac{N_z+i}{N_I-i+1}}\,\rho_{N_xN_y(N_z+k)}
\end{equation}
\end{widetext}

\bibliographystyle{apsrev4-1}
\bibliography{main}

\end{document}